\documentclass[11pt]{article}

\usepackage{geometry}                
\usepackage[parfill]{parskip}    
\usepackage{graphicx}
\usepackage{amsmath, amsthm, amsfonts}
\usepackage{lipsum}
\usepackage{newtxtext,newtxmath}

\newcommand{\sect}[1]{\setcounter{equation}{0}\section{#1}}

\usepackage{color}
     \usepackage[colorlinks]{hyperref}
\hypersetup{linkcolor=blue,%
citecolor=red,%
urlcolor=cyan}
\usepackage{url} 
\usepackage{epstopdf}
\title{Demkov--Fradkin tensor for curved harmonic oscillators}

\author{ 
\c{S}eng\"ul Kuru$^a$
\footnote{kuru@science.ankara.edu.tr, ORCID: \href{https://orcid.org/0000-0001-6380-280X}{0000-0001-6380-280X}}\,,
Javier Negro$^{b}$
\footnote{jnegro@fta.uva.es (Corresponding author), ORCID: 
\href{http://orcid.org/0000-0002-0847-6420}{0000-0002-0847-6420}}\,,
Sergio Salamanca$^b$
\footnote{sergio.salamanca@uva.es, ORCID: 
\href{https://orcid.org/0000-0003-0151-8373}{0000-0003-0151-8373}}   
\bigskip
\\
\noindent
$^a$\,Department of Physics, Faculty of Science, Ankara
University, 06100 Ankara,  T\"urkiye
\\ 
 \noindent
$^b$\,Departamento de F\'{\i}sica Te\'orica, At\'omica y
\'Optica, and IMUVA,\\ Universidad de Valladolid,  47011 Valladolid, Spain
}

\begin{document}

\maketitle

\begin{abstract}
In this work, we obtain the Demkov-Fradkin tensor of symmetries for the quantum curved harmonic oscillator in a  space with constant curvature given by a parameter $\kappa$. In order to construct this tensor we have firstly found a set of basic operators which satisfy the following conditions: i) their products give symmetries of the problem; in fact the Hamiltonian is a combination of such products; ii) they generate the space of eigenfunctions as well as the eigenvalues in an algebraic way; iii)
in the limit of zero curvature, they  come into the well known creation/annihilation operators of the flat oscillator. The appropriate products of such basic operators will produce the curved Demkov-Fradkin tensor. 
However, these basic operators do not satisfy Heisenberg commutators but close another Lie algebra. 
As a by-product, the classical  Demkov-Fradkin tensor for the classical curved harmonic oscillator has been obtained by the same method. 
The case of two dimensions has been worked out in detail:  the operators close a $so_\kappa(4)$ Lie algebra; the spectrum and eigenfunctions are explicitly solved in an algebraic way and in the classical case the trajectories have been computed. 
\end{abstract}

\sect{Introduction}
In classical and quantum mechanics to know symmetries of the systems is essential, because they give us information about the dynamics of these systems by means of related conserved quantities or some type of invariance. In general, systems may  have geometrical and dynamical 
symmetries, and sometimes may be of higher order in the momenta. It is well known  from Noether theorem \cite{castanos90,castanos92} that the continuous symmetries, such as translational, rotational or time independence, are related with the constants of motion (integrals of motion) of the system.  These constants of motion close an algebra, the symmetry algebra, that sometimes may be expressed as a Lie algebra. A system with $n$ degrees of freedom is said to be integrable when there are $n$ integrals of motion together with Hamiltonian commuting with  each other. If there exist $n-1$ additional independent integrals of motion commuting with the Hamiltonian, not necessarily  commuting with each other, the system is called maximally superintegrable \cite{winternitz13,post,evans90,evans08}. In classical mechanics we know from  Bertrand's theorem \cite{Bertrand} that the only two central systems, such that all the bounded trajectories are closed, are the harmonic oscillator (HO) and Kepler-Coulomb (KC) systems.  These two systems have enough symmetries, which include the Demkov--Fradkin (DF) tensor and  Runge-Lenz vector, respectively, what makes them maximally superintegrable.
It is known that the same happens for the HO and KC systems on constant curvature spaces: they are maximally superintegrable.

This paper is devoted to finding the Demkov--Fradkin tensor of the symmetries for the curved HO, that is, the HO in a constant curvature space. This tensor, in the case of three dimensions (3D), together with the angular momentum, include five independent symmetries corresponding to its maximal superintegrability.   There is a long list of contributions dealing with different aspects of symmetries and solutions for both classical and quantum curved HO. Some recent contributions, with many more references that can be found therein,  are \cite{
evans08,demkov59,fradkin65,suta19,ranada99,ranada02,ranada03,carinena08,carinena12,perelomov,ballesteros16,higgs79,kalnins97,kalnins01,tempesta01,salamanca23,ttw10}.

However, there is not a clearly established method to find the curved DF tensor of the curved HO in a way that resembles the flat construction, so that we intend to fill this point. Let us introduce the ingredients of of our approach. Consider in a  $n$-dimensional flat space  the harmonic oscillator $H$ of frequency $\omega$. Then, $H$ can be expressed as a sum of products of creation/annihilation operators corresponding to each dimension,
$H= \sum a^+_ia^-_i$. The operators $a^\pm_i$ are not symmetries, but they can be used to construct them, and these operators may be applied to generate the space of eigenfunctions. We can  say that these basic operators factorize in some way the HO Hamiltonian, in the sense  that the Hamiltonian can be expressed as linear combination of products. Then, what we are looking for is the kind of operators which play the same role of $a^\pm_i$ in the curved HO system, if they exist. In other words, we want to find the analogs of these basic operators such that:  i) A set of independent symmetries of the curved Hamiltonian can be obtained by means of products of such operators. ii) The Hamiltonian may be expressed as a linear combination of these products. iii) They should close an algebraic structure. iv) 
They must allow us to generate the space of eigenfunctions and to obtain the spectrum in an algebraic way.  v) In the limit $\kappa \to 0$ they should turn into  creation/annihilation operators of flat HO.

Our procedure, based on some previous works \cite{negro06,kuru08jpa} is original up to our knowledge. We will deal with this problem by computing  shift or intertwining operators which later will be identified to the basic operators we are looking for.
The meaning of shift or ladder operators originates from the factorization method (or Darboux transformations in a more general sense), which has been  applied 
mainly in the context of quantum mechanics, along a series of references  \cite{infeld,schrodinger40,salle91,cooper,junker,david}.  
Thus, our method consists in finding a kind of shift operators of first order. Although they are not symmetries, they are characterized by the fact that some of their products will supply them; in this sense these operators may be called pre-symmetries. The shift operators to be obtained in this work close a Lie algebra, which simplifies the solution to the problem.

We should remark that the method of using shift/ladder operators to find symmetries can be extended to classical systems by replacing the operators  by appropriate functions \cite{kuru08ann}. Indeed, we will find the classical DF tensor by  the same procedure, what will allow us to obtain the trajectories in a simple algebraic way. 

The organization of this work is as follows. We will start in section 2 by reviewing the Demkov-Fradkin tensor in flat space. We will work with creation/annihilation operators in Cartesian and polar coordinates as an initial point towards the operators to be found in curved spaces.
Next, in section~3 
we introduce a notation to state clearly what are the spaces which will substitute the Euclidean space $\mathbb R^3$ together with their ``free'' symmetry algebra. Afterwards, we will introduce the HO potential on these spaces along section~4. The basic operators and their algebra in geodesic parallel coordinates are obtained in section~5 and the corresponding symmetries are constructed. In a two dimensional space, a more appropriate basis of the algebra of operators in spherical coordinates is given in section~6, which simplifies the structure leading to symmetries and eigenstates. A short description of the curved classical HO system is included in section~7 and its trajectories in 2D are constructed. Some remarks are given in the final section.

%

\sect{The HO and Demkov--Fradkin tensor in flat space}

We will start by recalling briefly some general properties of DF tensor
in three dimensions, following the references \cite{demkov59, fradkin65},
where Cartesian coordinates were used.
%

\begin{itemize}
\item
{\it Cartesian coordinates} (in three dimensions)

The classical isotropic HO in ${\mathbb R}^3$  can be written, choosing appropriate units in the form
\begin{equation}\label{ho}
H_{\rm HO} = \sum_j-\partial_{jj} + \omega^2 {\bf r}^2
\end{equation}
where 
 ${\bf r}=(x_1,x_2,x_3)$ is the position vector, $\omega$ the frequency and $\partial_j =\frac{\partial}{\partial x_j}$, etc.. Due to the geometric spherical symmetry, the angular momentum components ${\bf J}:= (J_1,J_2,J_3)$ constitute three symmetries of the system:
\begin{equation}
{J_k} = \varepsilon_{kij}(-x_i\partial_j +x_j \partial_i),\qquad
L_k = iJ_k
\end{equation}
However, written in the form
\begin{equation}\label{ho}
H_{\rm HO} = \sum_{k=1}^{3} (-\partial_{kk} + \omega^2 x_k^2)
\end{equation}
we directly see that the HO Hamiltonian separates in Cartesian coordinates and each term $Q_{kk}$, 
\begin{equation}\label{qs}
Q_{kk}:=-\partial_{kk} + \omega^2 r_k^2 = a^+_k a^-_k + \omega\,, 
\qquad  
a^\pm_k=\mp \partial_k + \omega x_k\,, \qquad k=1,2,3
\end{equation}
is a one dimensional HO, which is a constant of motion. The factors 
 $a^\pm_k$ are creation/annihilation operators satisfying the Heisenberg commutators  
\begin{equation}\label{hopm}
[a^-_k, a^+_k] =   2\omega 
\end{equation}
Therefore, we have three symmetries in involution which correspond to the quantum number operators, 
\begin{equation}\label{hopm1}
N_k= a^+_k a^-_k = Q_{kk} ,\qquad k= 1,2,3 
\end{equation}
In particular, the Hamiltonian is a linear combination of these  involutive symmetries: 
\begin{equation}\label{hopm2}
H_{\rm HO}= N_1+N_2+N_3+3\omega
\end{equation}
The key point is that  the mixed products $Q_{ij}$ are also  symmetries:
\begin{equation}\label{hopm3}
Q_{ij}  = a^+_i a^-_j\,,\qquad  \{H_{\rm HO},Q_{ij}\} =0
\end{equation}
We can choose a set of five independent symmetries, as follows
\[
\{\ Q_{11}, Q_{22}, Q_{33}, Q_{12},Q_{23}\ \} = \{\ a_1^+a_1^-,\  a_2^+a_2^-,\  a_3^+a_3^-, \  a_1^+a_2^-,\  a_2^+a_3^-\}
\]
This means that the system is maximally superintegrable (in 3D $2n-1=5$). 
Since these symmetries may be non Hermitian, it is useful to construct Hermitian ones  $F_{jk}$ and  $D_{jk}$ by symmetric or antisymmetric linear combinations, which are second and first order, respectively:
\begin{equation}\label{hopm}
F_{jk}  := \frac12(a^+_j a^-_k + a^+_k a^-_j)= -\partial_j\partial_k + \omega ^2 x_j x_k \,,
\qquad  D_{jk}  := i\,\frac12(a^+_j a^-_k - a^+_k a^-_j)= -i\omega (x_j \partial_k-x_k \partial_j)= \omega L_{i}
\end{equation}
The axial vector
$D_{jk}$ is essentially the (Hermitian) angular momentum $
{\bf L}$, a symmetry that, for the classical HO, specifies the plane of the orbit.
The symmetric tensor $F_{jk}$ is called Demkov-Fradkin tensor which determines the orbit of the motion of the classical HO \cite{demkov59,fradkin65}. 
 






%

\item
\noindent
{\it Polar coordinates} (in two dimensions)

At this time it is convenient to consider a two-dimensional HO in polar coordinates, in order to check the consistency of the results that will be obtained later on, in sections 6 and 7.

If we restrict to the harmonic oscillator in the $xy$-plane, we make use of the symmetry in the angular momentum $J_{12}$ to describe the 2D HO in polar coordinates. Let us define the following operators
\begin{equation}\label{ABs}
\begin{array}{l}
A_+ := \frac12(a_1^+ + i a_2^+)\,,\qquad A_- := \frac12(a_1^- - i a_2^-)
\\[2.ex]
B_+ := \frac12(a_1^+ - i a_2^+)\,,\qquad B_- := \frac12(a_1^- + ia_2^-) 
\end{array}
\end{equation}
Their non zero commutation rules, together with $L_3$, are
\begin{equation}\label{ABcom}
[A_-,A_+] = \omega\,,\ \  [L_3,A_\pm] = \pm A_\pm\,,
\qquad\quad
[B_-,B_+] = -\omega\,,\ \  [L_3,B_\pm] = \pm B_\pm
\end{equation}
The HO Hamiltonian in this basis is expressed as
\begin{equation}\label{hABs}
H  = 4 A_+ A_-  - 2\omega L_3 + 2\omega
= 4 B_+ B_- + 2\omega L_3 - 2\omega
=  2(A_+ A_- + B_+ B_-) 
\end{equation}
Therefore,
\begin{equation}\label{comHAB}
[H,A_\pm] = \pm 2\omega A_\pm\qquad
[H,B_\pm] = \mp 2\omega B_\pm
\end{equation}

According to the previous formulas, we check that three independent symmetries are given by the operators 
\begin{equation}
S_1 = A_+ A_-\,,\quad S_2 = B_+ B_-\,,\quad
S_3 = B_+ A_+
\end{equation}
In order to get Hermitian symmetries we can make use of symmetric or antisymetric linear combinations,
\begin{equation}
F_{AA} = A_+ A_-\,,\quad F_{BB} = B_+ B_-\,,\quad
F_{AB} = B_+ A_+ + B_- A_-,\qquad
D_{AB} = i(B_+ A_+ - B_- A_-)
\end{equation}

Finally, for completeness we write the Hamiltonian operator $H$  in polar coordinates $(r,\phi)$. If we  take the wavefunctions $\Phi(r,\phi) = r^{-1/2} \Psi(r,\phi)$, then
\begin{equation}
H \Psi = \left(- \partial_{rr}- \frac{\partial_{\phi\phi} +1/4}{r^2} + \omega^2 r^2\right)\Psi = E \Psi
\end{equation}
while the operators $A_\pm,B_\pm$ take the form
\begin{equation}\label{ABs11}
A_+ = \frac12 \left( -\partial_r 
- \frac{i\partial_{\phi} + 1/2}{r}+ \omega r\right)e^{i\phi}\,,
\qquad
A_- =  \frac12\,e^{-i\phi} \left( \partial_r 
- \frac{i\partial_{\phi} +\,1/2}{r}+ \omega r\right)
\end{equation}
\begin{equation}\label{ABs22}
B_+ = \frac12 \left( -\partial_r 
+ \frac{i\partial_{\phi} +1/2}{r}+ \omega r\right)e^{i\phi}\,,
\qquad
B_- = \frac12 \,e^{-i\phi}\left(-\partial_r 
+ \frac{i\partial_{\phi} +1/2}{r}+ \omega r\right)
\end{equation}

\end{itemize}

Our motivation consists in the formulation of Demkov-Fradkin tensors in 
constant curvature spaces by introducing the analog of the ``basic'' operators $\{a_i^\pm\}$ in Cartesian coordinates, as well as those similar to $\{A_\pm,B_\pm\}$ in polar coordinates for the curved HO. 
We will show that from an algebraic point of view, the polar set will be most convenient in order to characterize the eigenfunctions and eigenvalues of the curved HO.
\sect{The free system on a constant curvature space} 

We will start 
by defining the relevant surfaces inside a four dimensional ambient space. 
Let us consider the surface $\Sigma_\kappa(x)$ in the ambient space $\mathbb R^4$ with coordinates  $(x_0,x_1,x_2,x_3)= (x_0,{\bf x})$ defined by
\begin{equation}\label{ksphere}
\Sigma_\kappa(x)=(x_0)^2 + \kappa
\, {\bf x}^2 =1
\end{equation} 
The parameter $\kappa$ being real is for the curvature, it has three representative values: $\kappa=1$ for the sphere, $\kappa=0$ for the 3D Euclidean space (with $x_0= 1$, in this case there is an additional spatial metric) and $\kappa= -1$ is for a hyperbolic space. The metric on the surface is induced from that of the ambient space. More details on this notation can be found in \cite{carinena08,carinena12,ballesteros06}.
The fields tangent to the surface $\Sigma_\kappa$ are
\begin{equation}\label{js}
J_{0i} = -x_0\partial_i+ \kappa\, x_i\partial_0,\qquad
J_{jk} = -x_j\partial_k +  x_k\partial_j,\qquad J_{jk} = -J_{kj} \quad\qquad j\neq k;\  i,j,k\neq 0
\end{equation}
Associated to these differential operators there are corresponding matrices:
\begin{equation}\label{jsm}
{\rm J}_{0i} = E_{i0}- \kappa E_{0i},\qquad 
{\rm J}_{jk} = E_{kj} -  E_{jk}
\end{equation}
where the basic matrices $E_{jk}$ have the unit in the  $j$-row and $k$-column position and zero otherwise.
The exponential of these matrix generators produce the following linear transformations leaving invariant the surface:
\begin{equation}\label{trans1}
e^{{\theta_{01}}{\rm J}_{01}}= \displaystyle \sum_{n=0}^{\infty}\frac{(\theta_{01}{\rm J}_{01})^{n}}{n!} 
=
\left(\begin{array}{cccc}
 \mathrm{C}_\kappa({\theta_{01}}) & -\kappa \mathrm{S}_\kappa({\theta_{01}})& 0& 0 \\
 \mathrm{S}_\kappa({\theta_{01}}) & \mathrm{C}_\kappa({\theta_{01}})&\  0\ &\  0 \  \\
 0 & 0& 1& 0   \\
 0 & 0& 0& 1   \\
\end{array}\right)
\end{equation}

\begin{equation}\label{trans2}
e^{{\theta_{12}}{\rm J}_{12}}= \displaystyle \sum_{n=0}^{\infty}\frac{(\theta_{12}{\rm J}_{12})^{n}}{n!} 
=
\left(\begin{array}{lccl}
 1 &0& 0& 0 \\
 0 &\cos \theta_{12}& -\sin \theta_{12}&\   \\
 0 & \sin \theta_{12}& \cos \theta_{12}& 0   \\
 0 & 0& 0& 1   \\
\end{array}\right)
\end{equation}

\begin{equation}\label{trans3}
e^{{\varphi_{23}}{\rm J}_{23}}= \displaystyle \sum_{n=0}^{\infty}\frac{(\varphi_{23}{\rm J}_{23})^{n}}{n!} 
=
\left(\begin{array}{cccc}
\  1\  &\ 0 \ & 0& 0 \\
 0 &1& 0& 0  \\
 0 &0& \cos \varphi_{23}& -\sin\varphi_{23}   \\
 0 & 0& \cos \varphi_{23}& \sin\varphi_{23}   \\
\end{array}\right)
\end{equation}
In order to deal simultaneously  with generic systems of any value of $\kappa$, we make use of the general  trigonometric functions following the notation described in \cite{ballesteros06}:
\begin{equation}\label{coord}
\mathrm{C}_\kappa(r) \displaystyle 
=
\left\{ \begin{array}{ll}
\cos{\sqrt{\kappa}r}, & \kappa>0;  \\
 1,& \kappa=0;    \\
  \cosh{\sqrt{-\kappa}r}, & \kappa<0;  \\
\end{array}\right.
\quad 
\mathrm{S}_\kappa(r) \displaystyle 
=
\left\{ \begin{array}{ll}
 \frac{1}{\sqrt{\kappa}} \sin{\sqrt{\kappa}r},& \kappa>0;  \\
 r  ,& \kappa=0;    \\
  \frac{1}{\sqrt{-\kappa}} \sinh{\sqrt{-\kappa}r} , & \kappa<0;  \\
\end{array}\right.
\end{equation}
The basic trigonometric relation is
\begin{equation}\label{normalizacion}
    \mathrm{C}^2_{\kappa}(\theta)+\kappa\,\mathrm{S}^2_{\kappa}(\theta)=1
\end{equation}
and the $\kappa$--tangent or $\kappa$--cotangent functions are
\begin{equation}\label{normalizacion}
    \mathrm{T}_{\kappa}(\theta) =\frac{ \mathrm{S}_{\kappa}(\theta)}{\mathrm{C}_{\kappa}(\theta)}\,,
    \qquad
    \mathrm{Cot}_{\kappa}(\theta) =\frac{ \mathrm{C}_{\kappa}(\theta)}{\mathrm{S}_{\kappa}(\theta)}
\end{equation}

Other properties can be found in \cite{ballesteros06}. These generators satisfy the following commutators (where only the nonvanishing ones  are given) 
\begin{equation}\label{comjs}
[J_{0i},J_{ik}] =  -J_{0k}\,\quad 
[J_{ij},J_{jk}] = -J_{ik}\,,\quad
[J_{0i},J_{0j}] = \kappa J_{ij}
\end{equation}
The quadratic Casimir coincides with the free Hamiltonian, and it is given by
\begin{equation}\label{cas}
{\cal H}^\kappa_0={\cal C}_\kappa = -\sum_{i} J_{0i}^2 - \kappa \sum_{i<j} J_{ij}^2 \,, \qquad i,j=1,2,3
\end{equation}
Thus, the symmetry algebra of the free Hamiltonian is spanned by basis 
$\{ \, J_{0i},\, J_{jk}\, \}$ of a Lie algebra that we call $so_\kappa(4)$.
For $\kappa=1$, $\Sigma_1$ is a sphere and the operators $\{ \, J_{0i},\, J_{jk}\, \}$  close $so(4)$; the value $\kappa=-1$ gives a hyperboloid $\Sigma_{-1}$ and the generators close $so(3,1)$. The case $\kappa \to 0$ can be obtained as a limit. The surface $\Sigma_0$ becomes the 3D flat space $x_0=1$, and the generators and commutation rules become
\begin{equation}\label{k0}
\begin{array}{l}
\qquad J_{0i}\  \to \ P_{i}=-\partial_i ,  
\qquad J_{jk}  \  \to \ J_{jk} \,, 
\\[2.ex]
[P_{i},J_{ik}] =  -P_{k}\,\qquad 
[J_{ij},J_{jk}] = -J_{ik}\,,\quad
[P_{i},P_{j}] = 0
\end{array}
\end{equation} 
where $P_i$ is the generator of translation in the $i$--direction. The resulting algebra is the Euclidean $iso(3)$  in three dimensions. In  $\kappa=0$ the Casimir becomes
\[
{\cal C}_0 = -\sum_{i} P_{i}^2 
\]

\sect{The quantum HO on a constant curvature space}

Next, we define the HO potential on the surface $\Sigma_\kappa$ 
(see \cite{ballesteros09}) in terms of the ambient coordinates. 
We will consider the following function on $\Sigma_\kappa$ as the definition of the HO potential:
\begin{equation}\label{pots}
 {V}^\kappa_{\rm HO}= {(\omega^2-\kappa^2/4)} \,
\frac{ {\bf x}^2 }{ {x_0}^2 }= 
{(\omega^2-\kappa^2/4)} \,
\frac{ {\bf x}^2 }{ 1-\kappa {\bf x}^2 } 
\end{equation}
The real parameter $\omega$ is for the frequency, while $\kappa$ refers to the curvature. For $\kappa>0$ the radius is finite, ${\bf x}^2<1/\kappa$.    
We will see later that 
the relation of frequency $\omega$ and curvature $\kappa$ determine what representations describe the states of the quantum system. 

We will also take into account another auxiliar potential, which differ from the previous one in a constant:
\begin{equation}\label{pots2}
\bar{V}^\kappa_{\rm HO}=\frac{\omega^2-\kappa^2/4}{\kappa {x_0}^2}\,,
\qquad 
\bar{V}^\kappa_{\rm HO} = {V}^\kappa_{\rm HO} + 
\frac{\omega^2-\kappa^2/4}{\kappa}
\end{equation}
Remark that in the limit $\kappa \to 0$, the potential ${V}^\kappa_{\rm HO}$ has a well defined limit towards that of the flat oscillator: ${V}^0_{\rm HO}=\omega ^2 {\bf x}^2$. However, $\bar{V}^\kappa_{\rm HO}$ diverges, so the limit has to be computed after including the constant $\frac{\omega^2-\kappa^2/4}{\kappa}$. Both potentials give rise to the same symmetries, but we will choose $\bar{V}^\kappa_{\rm HO}$ in some computations because it has simpler factorization properties. Both potentials (\ref{pots2}) are singular on the equator $x_0=0$.


We define the HO Hamiltonians corresponding to these two potentials, by
\begin{equation}\label{hhh}
\mathcal{H}_{\rm HO}^{\kappa} = \mathcal{H}^{\kappa}_0 + 
V^{\kappa}_{\rm HO},\qquad
\bar{\mathcal{H}}_{\rm HO}^{\kappa} = \mathcal{H}^{\kappa}_0 +\bar{V}^{\kappa}_{\rm HO}
\end{equation}
or, explicitly,
\begin{equation}\label{hhh2}
\mathcal{H}_{\rm HO}^{\kappa} = -\sum_{i} J_{0i}^2 
+ {(\omega^2-\kappa^2/4)} \,
\frac{ {\bf x}^2 }{ {x_0}^2 }
- \kappa \sum_{i<j} J_{ij}^2 \,,
\qquad
\bar{\mathcal{H}}_{\rm HO}^{\kappa} =-\sum_{i} J_{0i}^2 
+\frac{(\omega^2-\kappa^2/4)}{\kappa {x_0}^2}
- \kappa \sum_{i<j} J_{ij}^2  
\end{equation}

If necessary, we will add the superindex $\omega$ to specify the frequency of the potentials in these Hamiltonians: $\mathcal{H}_{\rm HO}^{\kappa,\omega}$ and $\bar{\mathcal{H}}_{\rm HO}^{\kappa,\omega}$. We will always assume that 
$\bar{\mathcal{H}}_{\rm HO}^{\kappa,\omega}$ is an auxiliary Hamiltonian.

In a similar way to the flat oscillator, we may use different coordinate systems of the surface in order to separate variables: spherical and the analog of Cartesian, the geodesic parallel coordinates.
\begin{itemize}
\item {\it Spherical coordinates}

Since the operators $J_{ij}$ corresponding to  spatial rotations are symmetries of the HO system (\ref{hhh2}), it will be separable in spherical coordinates.   An spherical coordinate chart corresponding to the coordinates $(\theta_{01}, \theta_{12}, \phi_{23})$ is obtained as follows:
\begin{equation}\label{coord4s0}
x_0 = \mathrm{C}_{\kappa}(\theta_{01})\,, \quad
x_1=  \mathrm{S}_{\kappa} (\theta_{01}) \cos \theta_{12}\,,\quad  
x_2= \mathrm{S}_{\kappa} (\theta_{01}) \sin \theta_{12} \cos\phi_{23}\,,\quad
x_3= \mathrm{S}_{\kappa} (\theta_{01})\sin \theta_{12} \sin \phi_{23}
\end{equation}
where, the angle $\theta_{01}$  would correspond to a rotation by $J_{01}$ in the $x_0x_1$--plane  of amplitude $\theta_{01}$ (leaving unaltered the rest of coordinates). The same can be said for the angles $\theta_{12}$ and $\phi_{23}$.
The  expressions for the HO potential, ${V}^\kappa_{\rm HO}$, and the auxiliary potential,   $\bar{V}^\kappa_{\rm HO}$, on the surface $\Sigma_\kappa$ in this set of spherical coordinates take the form
\begin{equation}\label{pots3}
{V}^\kappa_{\rm HO}= {(\omega^2-\kappa^2/4)}
\mathrm{T}^2_{\kappa} (\theta_{01})\,,
\qquad 
\bar{V}^\kappa_{\rm HO} 
=  \frac{\omega^2-\kappa^2/4}{ \kappa \,{\mathrm{C}^2_{\kappa}} (\theta_{01})}
\end{equation}
The expression for $\bar{\mathcal{H}}_{\rm HO}^{\kappa}$ in the set (\ref{coord4s0}) of spherical coordinates is
\begin{equation}\label{hamiltonianC}
\begin{array}{l}
\displaystyle
\bar{\mathcal{H}}_{\rm HO}^{\kappa}= 
-\partial^2_{\theta_{01}}
-\frac{2}{\mathrm{T}_{\kappa}(\theta_{01})}\partial_{\theta_{01}}
+\frac{\omega^2-\kappa^2/4}{\kappa\,\mathrm{C}^2_{\kappa}(\theta_{01})}
+\frac{1}{\mathrm{S}^2_{\kappa}(\theta_{01})}
\Big(-\partial^2_{\theta_{12}}
-\frac{1}{\tan \theta_{12} }\partial_{\theta_{12}}
+\frac{1}{\sin^2 \theta_{12}}
\big(-\partial^2_{\phi_{23}}  \big)\Big) 
\end{array}
\end{equation}
From this expression it is clear that  the (square of)  rotation generator $J_1=J_{23}$, 
\begin{equation}
I_{23}=\partial^2_{\phi_{23}} = (-x_2 \partial_3 + x_3 \partial_2)^2=J_{23}^2
\end{equation} 
is a symmetry of the HO Hamiltonians (\ref{hhh}):
\begin{equation}
[I_{23},\bar{\mathcal{H}}_{\rm HO}^{\kappa}] = [I_{23},{\mathcal{H}}_{\rm HO}^{\kappa}]= 0
\end{equation} 
 In the same way, all the `spacial' rotations 
\begin{equation}\label{ies}
I_{12},\quad I_{23},\quad I_{13};\qquad 
I_{jk}=(-x_j \partial_k + x_k \partial_j)^2= J_{jk}^2, \qquad i,j= x,y,z
\end{equation} 
are symmetries.
This is consistent with our starting point, since the potentials 
$\bar{ V}^\kappa_{\rm HO}$ and ${V}^\kappa_{\rm HO}$ are invariant under `spatial rotations' (not involving the coordinate $x_0$).

\item {\it Geodesic parallel coordinates}

In order to make explicit other symmetries we adopt  different coordinate systems,
involving the rotations $J_{0i}$ which are not symmetries.
Consider a set of geodesic parallel coordinates \cite{ballesteros06}, $(\vartheta_{01}, \vartheta_{02}, \varphi_{03})$ parametrising the surface in the following way,
\begin{equation}\label{coord4s}
x_1=  \mathrm{S}_{\kappa}(\vartheta_{01})\,,\quad
x_2=  \mathrm{C}_\kappa(\vartheta_{01})\mathrm{S}_\kappa(\vartheta_{02})\,,  \quad   
x_3= \mathrm{C}_\kappa(\vartheta_{01})\mathrm{C}_\kappa(\vartheta_{02})\mathrm{S}_\kappa(\varphi_{03}) \,,\quad
x_0 = \mathrm{C}_\kappa(\vartheta_{01})\mathrm{C}_\kappa(\vartheta_{02})\mathrm{C}_\kappa(\varphi_{03})                               
\end{equation}
Then, we will get a similar expression for the Hamiltonian with the new angles:
\begin{equation}\label{hamiltonianC}
\begin{array}{l}
\displaystyle
\bar{\mathcal{H}}_{\rm HO}^{\kappa}=
-\partial^2_{\vartheta_{01}}
-\frac{2}{\mathrm{T}_{\kappa}(\vartheta_{01})}\partial_{\vartheta_{01}}
+\frac{1}{\mathrm{C}_{\kappa}^2(\vartheta_{01})} 
\Big(-\partial^2_{\vartheta_{02}}
-\frac{1}{\mathrm{T}_{\kappa}(\vartheta_{02})}\partial_{\vartheta_{02}}
+\frac{1}{ \mathrm{C}_{\kappa}^2(\vartheta_{02}) }
\big(-\partial^2_{\varphi_{03}} 
+\frac{\omega^2-\kappa^2/4}{\kappa\,\mathrm{C}_{\kappa}^2(\varphi_{03})}\big)\Big)
\end{array}
%
\end{equation}
However, from this expression we get that  the operator
\begin{equation}\label{sym42}
\bar{I}_{03}^{\kappa,\omega}= -\partial^2_{\varphi_{03}} 
+\frac{\omega^2-\kappa^2/4}{\kappa\,\mathrm{C}_{\kappa}^2(\varphi_{03})}
\end{equation}
or
\begin{equation}\label{sym42b}
{I}_{03}^{\kappa,\omega}= -\partial^2_{\varphi_{03}} 
+{(\omega^2-\kappa^2/4)}{ \mathrm{T}_{\kappa}^2(\varphi_{03})} 
\end{equation}
will constitute another symmetry of any of the Hamiltonians $\bar{\mathcal{H}}_{\rm HO}^{\kappa}$ or ${\mathcal{H}}_{\rm HO}^{\kappa}$.  In fact, we will have three symmetries by making use of the geodesic angles $\varphi_{01},\varphi_{02},\varphi_{03}$
(each one belong to a different set of parallel coordinates),
\begin{equation}\label{i0k0}
{I}_{01}^{\kappa,\omega}\,,\quad  {I}_{02}^{\kappa,\omega}\,, \quad  {I}_{03}^{\kappa,\omega}
\end{equation}
In conclusion, up to now, we have six symmetries of the curved HO (\ref{ies}) and (\ref{i0k0}):
\[
I_{12}\,,\quad I_{23},\quad I_{13},\quad {I}_{01}^{\kappa,\omega}\,,\quad  {I}_{02}^{\kappa,\omega}\,, \quad  {I}_{03}^{\kappa,\omega}
\] 
They explain the separation of the Hamiltonian in the above sets of coordinates. Although these symmetries were obtained by means of particular coordinate systems, they all can be expressed in a unique coordinate system. We will make use of ambient coordinates to write all of them.  

We can express the total HO Hamiltonian (\ref{hhh2}) in terms of these symmetries,
\begin{equation}\label{hhh3}
\mathcal{H}_{\rm HO}^{\kappa} = \sum_{i} {I}_{0i}^{\kappa,\omega} \,
- \kappa \sum_{j<k} I_{jk} 
\end{equation}

\end{itemize}

\sect{Shift operators and generalized Demkov--Fradkin tensor}

Next, we will show how  shift (or intertwining) operators will be relevant. Let us take for instance, the symmetry operator (\ref{sym42}). Since it has the form of a one-parameter P\"oschl-Teller Hamiltonian \cite{kuru9},  it can be  factorized in the usual form:
\begin{equation}\label{sym42c}
\bar{I}_{03}^{\kappa,\omega}=a^+_{03}(\omega) a^-_{03}(\omega)+
\mu(\omega)\,,\qquad
\bar{I}_{03}^{\kappa,\omega+\kappa}=a^-_{03}(\omega) a^+_{03}(\omega)+
\mu(\omega)\,,
\qquad 
 \mu(\omega)=  \frac{(\omega+\kappa/2)^2}{\kappa}
\end{equation}
with the factorization operators (in ambient space coordinates) given by
\begin{equation}\label{aas}
a^\mp_{03}(\omega)=\pm\partial_{\varphi_{03}} +{(\omega+\kappa/2)}\mathrm{T}_{\kappa}(\varphi_{03}),\qquad
\partial_{\varphi_{03}}= x_0\partial_3 - \kappa\, x_3\partial_0\,,
 \qquad 
 \mathrm{T}_{\kappa}(\varphi_{03})=\frac{x_3}{x_0}
\end{equation}
The operators $a^\pm_{03}(\omega)$ of (\ref{aas}) are shift operators (sometimes we will use the more complete notation $a^\pm_{03}(\kappa,\omega)$), changing the parameter $\omega$
of $\bar I_{03}^{\kappa,\omega}$ in  $\kappa$-steps by means of an intertwining relation:
\begin{equation}\label{shift}
a^-_{03}(\omega)\, \bar I_{03}^{\kappa,\omega}
= \bar I_{03}^{\kappa,\omega+\kappa}\, a^-_{03}(\omega),
\qquad
a^+_{03}(\omega)\, \bar I_{03}^{\kappa,\omega+\kappa}
= \bar I_{03}^{\kappa,\omega}\, a^+_{03}(\omega)
\end{equation}

Taking into account the expression of Hamiltonian (\ref{hamiltonianC}), these are also
shift operators for $\bar{\mathcal{H}}_{\rm HO}^{\kappa,\omega}$, where we have added the superindex $\omega$ because it is modified under the action of the shift operators:
\begin{equation}\label{key}
 a^-_{03}(\omega)\,\bar{\mathcal{H}}_{\rm HO}^{\kappa,\omega}
= 
\bar{\mathcal{H}}_{\rm HO}^{\kappa,\omega+\kappa}\, a^-_{03}(\omega)
\end{equation}


We can also factorize  the equivalent, well behaved, symmetries ${I}_{03}^{\kappa,\omega}$
\begin{equation}\label{sym42d}
{I}_{03}^{\kappa,\omega}=a^+_{03}(\omega) a^-_{03}(\omega) +
(\omega+\kappa/2)\,,\qquad
{I}_{03}^{\kappa,\omega+\kappa}=a^-_{03}(\omega) a^+_{03}(\omega) -
(\omega+\kappa/2)
\end{equation}
The intertwining relations of the Hamiltonian ${\mathcal{H}}_{\rm HO}^{\kappa,\omega}$ can be easily obtained from (\ref{key}) or (\ref{sym42d}),
\begin{equation}\label{key2}
a^-_{03}(\omega)\,\mathcal{H}_{\rm HO}^{\kappa,\omega} 
= 
\left(\mathcal{H}_{\rm HO}^{\kappa,\omega+\kappa} +2(\omega + \kappa/2)\right) 
a^-_{03}(\omega)
\end{equation}
Therefore, with respect to the Hamiltonians ${\mathcal{H}}_{\rm HO}^{\kappa,\omega}$, the operators
$a^{-}_{03}(\omega)$ are mixed shift-ladder operators (they also change in $2(\omega + \kappa/2)$ the eingevalue). It is in this sense that the factorization properties of
$\bar{\mathcal{H}}_{\rm HO}^{\kappa,\omega}$ are simpler.

From the key relation (\ref{key}), we prove easily the form of the symmetries
(\ref{sym42c}), where we adopt the following notation, 
\begin{equation}\label{sym4jb}
Q_{jj}^{\kappa,\omega}:=a^+_{0j}(\omega) a^-_{0j}(\omega)
\end{equation}
In fact, we  can prove that there are additional symmetries by means of mixed products of 
shift operators:
\begin{equation}\label{sym4jc}
Q_{ij}^{\kappa,\omega}:=a^+_{0i}(\omega) a^-_{0j}(\omega),\quad i,j=1,2,3
\end{equation}

%
This is quite similar to the form of flat symmetries given previously in (\ref{hopm1}) and (\ref{hopm3}).
We have the following limits when $\kappa \to 0$:
\[
a^\pm_{0i}(\kappa,\omega)\ \to \ \mp\partial_i+ \omega x_i = a_i^\pm
\]
corresponding to the creation/annihilation operators of flat HO (\ref{aas}) in Cartesian coordinates.

Remark that

i) $Q_{ij}$ are symmetries for both $\bar{\mathcal{H}}_{\rm HO}^{\kappa,\omega}$ and $\mathcal{H}_{\rm HO}^{\kappa,\omega}$\,.

ii) $Q_{ij}$ have a well defined   limit in $\kappa\to 0 $, $x_0\to 1$ to the symmetries of flat HO.

In the following, we will simplify  the notation of shift operators by eliminating the index $0$, replacing $a^\pm_{0j}(\kappa,\omega)$ by $a^\pm_{j}(\kappa,\omega)$. 
Let us write the commutators of these basic operators (completed with those with $J_{ij}$ given in (\ref{comjs})):
\begin{equation}\label{aes0}
\begin{array}{l} 
\displaystyle
[a^-_{i}(\kappa,\omega),a^+_{i}(\kappa,\omega)]= 2 \omega 
\\[2.ex]
\displaystyle
[a^-_{i}(\kappa,\omega),a^+_{j}(\kappa,\omega)]=
[a^+_{i}(\kappa,\omega),a^-_{j}(\kappa,\omega)]=-2 \kappa J_{ij}
\\[2.ex]
\displaystyle
[a^+_{i}(\kappa,\omega),a^+_{j}(\kappa,\omega)]= [a^-_{i}(\kappa,\omega),a^-_{j}(\kappa,\omega)]=0
\\[2.ex]
\displaystyle
[J_{ij},a^\pm_{j}(\kappa,\omega)]=
 - a^\pm_{i}(\kappa,\omega)
\end{array}
\end{equation}
We should perform these commutation rules having in mind the action of $a^\pm(\omega)$ given in (\ref{key})-(\ref{key2}). For instance, the first commutator
in the list (\ref{aes0}), should be computed in the form
\begin{equation}\label{ws}
[a^-_{i}(\kappa,\omega),a^+_{i}(\kappa,\omega)]:= 
a^-_{i}(\kappa,\omega-\kappa),a^+_{i}(\kappa,\omega-\kappa)-
a^+_{i}(\kappa,\omega),a^-_{i}(\kappa,\omega) =  2 \omega 
\end{equation}
Remark that the commutation rules (\ref{aes0}) are different to the standard commutators of flat creation/annihilation operators $a^\pm_i$ of section~2. In this case the operators $a^\pm_{i}(\kappa,\omega)$
do not close a subalgebra.

The explicit expression of the (two dimensional) HO Hamiltonian
(\ref{hhh2}) in terms of shift operators is
\begin{equation}\label{hhh2b}
\mathcal{H}_{\rm HO}^{\kappa,\omega} = \sum_{i} a^+_i(\kappa,\omega) a^-_i(\kappa,\omega) + 2(\omega + \kappa/2) \,
- \kappa \sum_{j<k} I_{jk} 
\end{equation} 
Making use of the above commutation rules, one can check that $Q_{ij}$ as defined in  (\ref{sym4jc}) are indeed symmetries of this Hamiltonian.

Since the effect of the operators $a_i^\pm(\omega)$ is to change the frequency $\omega$, we may apply the replacement 
\[
\omega \to -i \partial_\zeta
\]
together with
\begin{equation}\label{zetas}
\begin{array}{l}
a^-_{3}(\kappa,\omega)=\partial_{\varphi_{03}} +{(\omega+\kappa/2)}\mathrm{T}_{\kappa}(\varphi_{03})
\ \longrightarrow \ 
a^-_{3}(\kappa,\zeta)=e^{i\kappa \zeta} \left(\partial_{\varphi_{03}} +{(-i \partial_\zeta+\kappa/2)}\mathrm{T}_{\kappa}(\varphi_{03}\right)
\\[2.ex]
a^+_{3}(\kappa,\omega)=-\partial_{\varphi_{03}} +{(\omega+\kappa/2)}\mathrm{T}_{\kappa}(\varphi_{03})
\ \longrightarrow \ 
a^+_{3}(\kappa,\zeta)= \left(\pm\partial_{\varphi_{03}} +{(-i \partial_\zeta+\kappa/2)}\mathrm{T}_{\kappa}(\varphi_{03}\right) e^{-i\kappa \zeta}
\end{array}
\end{equation}
We will assume that the operators act on the wavefunctions of the form
\begin{equation}\label{psis2}
\Psi^\kappa_\omega (x_0,{\bf x},\zeta) = 
e^{i \omega \zeta}
\psi_{\omega}^\kappa(x_0,{\bf x})
\end{equation}

After this change, we obtain ``true commutators'' of operators:
\begin{equation}\label{aes}
\begin{array}{l} 
\displaystyle
[a^-_{i}(\kappa,\zeta),a^+_{i}(\kappa,\zeta)]= -2i k \partial_\zeta  
\\[2.ex]
\displaystyle
[a^-_{i}(\kappa,\zeta),a^+_{j}(\kappa,\zeta)]=
[a^+_{i}(\kappa,\omega),a^-_{j}(\kappa,\zeta)]=-2 \kappa J_{ij}
\\[2.ex]
\displaystyle
[-i \partial_\zeta\,,a^\pm_{i}(\kappa,\zeta)]= \mp \kappa a^\pm_{i}(\kappa,\zeta)
\\[2.ex]
\displaystyle
[a^+_{i}(\kappa,\zeta),a^+_{j}(\kappa,\zeta)]= [a^-_{i}(\kappa,\zeta),a^-_{j}(\kappa,\zeta)]=0
\\[2.ex]
\displaystyle
[J_{ij},a^\pm_{j}(\kappa,\zeta)]=
 - a^\pm_{i}(\kappa,\zeta)
\end{array}
\end{equation}
Notice that these commutators of the basis 
$\{ a^\pm_{i}, J_{ij}, \partial_\zeta \}$ close a true Lie algebra.  We will see in detail later 
for two space dimensions in the ambient space
$(x_0,x_1,x_2)$, what algebra is and its applications.  

Since the Hamiltonian (\ref{hhh2b}) (and the symmetries) depends on products
$a_i^+ a_j^-$, it will depend on $\zeta$ only through the derivative $-i\partial_\zeta$, while the variable $\zeta$ is ignorable, so that this operator may be replaced by the constant $\omega$ on each physical eigenspace, which shows that there is an equivalence in the description in terms of $\omega$ or in  $\zeta$.

We want to discover the properties of the HO system by means of the intertwining
operators $a^\pm_k(\kappa,\omega)$. One advantage is that these operators close a Lie algebra. The symmetries as well as the Hamiltonian will be expressed in terms of such operators.
The eigenvalues and corresponding eigenfunctions of the HO belong to 
the support space of representations of the intertwining Lie algebra. We call these basic operators geodesic parallel or simply parallel shift operators, since they where obtained from factorization of symmetries in geodesic parallel coordinates.

\subsection{Symmetric and antisymmetric tensors of symmetries of curved HO}
%
%
Since the symmetries (\ref{sym4jc}) are non Hermitian, it may be useful to obtain Hermitian ones by symmetrization,
\begin{equation}\label{fdc}
F_{ij}^{\kappa}:= \frac12 \Big(a^+_{j}(\kappa,\omega) a^-_{i}(\kappa,\omega) + a^+_{i}(\kappa,\omega) a^-_{j}(\kappa,\omega)\Big),\quad
D_{jk}^{\kappa}:= \frac1{2 i} \Big(a^+_{j}(\kappa,\omega) a^-_{k}(\kappa,\omega) - a^+_{k}(\kappa,\omega) a^-_{j}(\kappa,\omega)\Big)
\end{equation}
Explicitly
\begin{equation}\label{aij}
F_{ij}^{\kappa }=
\frac{ x_i x_j (\omega^2-\kappa^2/4)} {x_0^2} 
    -\frac12\,(J_{0i}J_{0j}+J_{0j}J_{0i})\,, \qquad 
D_{ij}^{\kappa}= -(\omega+\kappa) \, J_{ij}
      \end{equation}
Its components are symmetries
 \begin{equation}
  [ F_{ij}^{\kappa},{\mathcal{H}}_{\rm HO}^{\kappa,\omega}]=0
   \end{equation}
and they satisfy
    \begin{equation}
 \sum_j F_{ik}^{\kappa} J_{k}=\sum_i J_{k} F_{ik}^{\kappa}=\kappa J_{i}
  \end{equation}
where as usual, $J_k=\epsilon_{kij}J_{ij}$ are angular momentum components. 
From (\ref{fdc}), the HO Hamiltonian (\ref{hhh2}) can be rewritten as:
 \begin{equation}\label{haes}
    {\mathcal{H}}_{\rm HO}^{\kappa,\omega} =
    \sum_i 
    F_{ii}^{\kappa} +2(\omega+\kappa/2)
    + \kappa \sum_{i<j} J_{ij}^2  
   \end{equation}

The tensor $F_{ij}^{\kappa }$ is the analog of Demkov--Fradkin tensor of symmetries for the curved HO.
In the limit $\kappa \to 0$, $x_0 \to 1$, we get the flat Demkov--Fradkin tensor \cite{fradkin65}:  
   \begin{equation}
   F^{\kappa=0}_{ij}= {\omega^2 x_i x_j} -  \partial_{i}\partial_{j}
   \end{equation}

Notice that there are differences of this expression for the curved HO Hamiltonian with respect to the flat DF tensor and HO Hamiltonian. There is a term in $\kappa$ which dissappear in the limit $\kappa \to 0$.  

Remark that the expressions of the basic operators $a^\pm_{j}(\kappa,\omega)$, as well as the tensors
$F_{ij}^\kappa$, $D_{ij}^\kappa$ are extensible to any dimension $n$, so that
they are not restricted to 3D.

\sect{Symmetries in polar coordinates}

\subsection{Shift operators in polar basis}

Next, we will write the shift operators in another basis. Hereafter, we will restrict to two dimensions where $i,j=1,2$) to simplify the algebraic structure. The same change of basis can be carry out in three dimensions, but here we will focus on the simplest 2D case.  This new basis is adapted to spherical variables which will be introduced in the following. Let us define this new basis as follows 
\begin{equation}\label{aijps}
\begin{array}{l}
\displaystyle
A^+_{p}(\kappa,\omega)= 
\frac12\Big(a^+_{1}(\kappa,\omega)\ +  i\,a^+_{2}(\kappa,\omega)\Big)
\\[2ex]
\displaystyle
A^-_{m}(\kappa,\omega)= 
\frac12\Big(a^-_{1}(\kappa,\omega)\ - i\,a^-_{2}(\kappa,\omega)\Big)
\\[2ex]
\displaystyle
A_3(\kappa,\omega)=\frac12\Big(-\omega + 
i\, \kappa J_3 \Big)
\\[2ex]
\displaystyle
B^+_{m}(\kappa,\omega)= 
\frac12\Big(a^+_{1}(\kappa,\omega) \ - i\,a^+_{2}(\kappa,\omega)\Big)
\\[2ex]
\displaystyle
B^-_{p}(\kappa,\omega)= 
\frac12\Big(a^-_{1}(\kappa,\omega) \ + i\,a^-_{2}(\kappa,\omega)\Big)
\\[2ex]
\displaystyle
B_3(\kappa,\omega)=\frac12\Big(  \omega + 
i\, \kappa J_3 \Big)
\end{array}
\end{equation}

The commutation rules (having in mind that $\omega$ must change according to (\ref{ws})) are as follows
\begin{equation}\label{sop1}
\displaystyle
[A^+_{p},A^-_{m}]= 2 A_3,
\quad
[A_3,A^+_{p}]= \kappa A^+_{p},\quad [A_3,A^-_{m}]= -\kappa A^-_{m}
\end{equation}
\begin{equation}\label{sop2}
\displaystyle
[B^+_{m},B^-_{p}]= -2 B_3,
\quad
[B_3,B^+_{m}]= -\kappa B^+_{m},\quad [B_3,B^-_{p}]= \kappa B^-_{p}
\end{equation}
The  $B's$ operators commute with $A's$.
All this means that we have a direct sum of two $so_{\kappa}(3)$ copies.
Hereafter, we will simplify the notation keeping one index according to the following convention,
\begin{equation}\label{not}
A_+:=A^+_{p},\quad A_-:=A^-_{m};\qquad B_+:=B^-_{p},\quad B_-:=B^+_{m}
\end{equation}
In this notation, we have a direct sum subalgebra,
\[
\langle A_\pm, A_3\rangle\oplus \langle B_\pm,  B_3 \rangle = 
so_{\kappa}(3)\oplus so_\kappa(3)
\]
If ${\kappa}=1$ this algebra is $so(3)\oplus so(3)\approx so(4)$; for ${\kappa}=-1$ this is $so(2,1)\oplus so(2,1)\approx so(2,2)$ \cite{negro24}.
For ${\kappa}=0$ we have the flat oscillator in two dimensions which consist in  the direct sum of two independent Heisenberg algebras, $h(1)\oplus h(1)$.
We must also include the rotations in the perpendicular axis $J_3$ act on these algebras in a semidirect way. Let us call $L_3$ to the Hermitian version of $J_3$ 
and $\ell$ to its eigenvalues,
\[
L_3:= i J_3= -i \partial_\phi\,,\qquad L_3\,\Psi_{\ell} = \ell\,\Psi_{\ell}
\]
Then, the commutation with the Hermitian generator $L_3$ is as follows
\begin{equation}\label{sop3}
[L_3,A_\pm]= \pm A_\pm,\qquad [L_3,B_\pm]= \pm B_\pm
\end{equation}

We have two Casimir operators corresponding to each subalgebra (in $\kappa=0$ we have to perform a limit $\kappa \to 0$, as we will see later):
\begin{equation}\label{cas}
\begin{array}{l}
\displaystyle
{\cal C}_1= \frac{4}{\kappa}
\left(\kappa\, A_-(\omega -\kappa)A_+(\omega -\kappa)
+A_3(\kappa,\omega)(A_3(\kappa,\omega)+\kappa)\right)
\\[2.ex]
\qquad \qquad=
\frac{4}{\kappa}
\left(\kappa\, A_+(\omega )A_-(\omega )
+A_3(\kappa,\omega)(A_3(\kappa,\omega)-\kappa)\right)
\\[2ex]
\displaystyle
{\cal C}_2= 
\frac{4}{\kappa}
\left(\kappa\, B_-(\omega)B_+(\omega )
+B_3(\kappa,\omega)(B_3(\kappa,\omega)+\kappa)\right)
\\[2.ex]
\qquad \qquad = 
\frac{4}{\kappa}
\left(\kappa\, B_+(\omega-\kappa)B_-(\omega-\kappa )
+B_3(\kappa,\omega)(B_3(\kappa,\omega)-\kappa)\right)
\end{array}
\end{equation}

Next, we will compute the differential realization of these operators in spherical coordinates $(\theta,\phi)$, introduced earlier (\ref{coord4s0}), with a slight change of the notation for the angles due to the restriction of coordinates to $(x_0,x_1,x_2)$,
\begin{equation}\label{coord4s0b}
x_0 = \mathrm{C}_{\kappa}(\theta)\,, \quad
x_1=  \mathrm{S}_{\kappa} (\theta ) \cos \phi\,,\quad  
x_2= \mathrm{S}_{\kappa} (\theta) \sin \phi 
\end{equation}
We can express all the operators and the commutation rules in terms of a differential realization as follows. 
As before, we use of a new variable $\zeta$ and its conjugate $-i\partial_\zeta$, instead of $\omega$.
Therefore, we will assume that the operators act on wavefunctions of the form
\begin{equation}\label{psis}
\Psi^\kappa_{\omega,\ell} (\theta,\phi,\zeta) = 
e^{i\ell \phi + i \omega \zeta}
\psi^\kappa_{\omega,\ell}(\theta)
\end{equation}
Then, we obtain the following differential realization for these generators
\begin{equation}\label{abs}
\begin{array}{l}
\displaystyle
A_+= \frac12  \left(-\partial_\theta 
-\frac{1/2+i \partial_\phi}{{\rm T}_\kappa(\theta)} + (-i\partial_\zeta+\kappa/2) {\rm T}_\kappa(\theta)\right)e^{i\phi}e^{-i\zeta \kappa}
\\[2.ex]
\displaystyle
A_-= \frac12 e^{-i\phi}e^{i\zeta \kappa} \left(\partial_\theta 
-\frac{1/2+i \partial_\phi}{{\rm T}_\kappa(\theta)} + (-i\partial_\zeta+\kappa/2){\rm T}_\kappa(\theta)\right)
\\[2.ex]
\displaystyle
A_3 =\frac12\Big(i \partial_\zeta - 
i\, \kappa \partial_\phi \Big)
\\[2ex]
\displaystyle
B_-= \frac12 e^{-i\phi} \left(-\partial_\theta 
+\frac{1/2+i \partial_\phi}{{\rm T}_\kappa(\theta)} + (-i\partial_\zeta+\kappa/2) {\rm T}_\kappa(\theta)\right) e^{-i\zeta \kappa}
\\[2.ex]
\displaystyle
B_+= \frac12 e^{i\zeta \kappa}\left(\partial_\theta 
+\frac{1/2+i\partial_\phi}{{\rm T}_\kappa(\theta)} + (-i\partial_\zeta+\kappa/2) {\rm T}_\kappa(\theta)\right) e^{i\phi}
\\[2.ex]
\displaystyle
B_3=\frac12\Big(-i \partial_\zeta - 
i\, \kappa \partial_\phi \Big)
\end{array}
\end{equation}
Remark that they are the well known factoriazations of the two-parametric P\"ochl-Teller Hamiltonian \cite{kuru12}. Therefore, we have a basis related to the variables $\phi$ and $\zeta$, corresponding to two generators $L_3=-i\partial_\phi$
and $\Omega:= -i\partial_\zeta$, which commute:
\[
[\Omega, L_3] =0
\]
Then, the rest of generators are determined by their eigenvalues
(\ref{sop1}), (\ref{sop2}), (\ref{sop3}) together with
\begin{equation}\label{sop4}
[\Omega,A_\pm]= \mp \kappa A_\pm,\qquad [\Omega,B_\pm]= \pm \kappa B_\pm
\end{equation}
This is what we call spherical basis, as opposed to the previous parallel basis.  
Let us remark some properties of the Casimir operators in this basis.
\begin{itemize}
\item[(i)] 

In the actual differential realization the Casimir operators ${\cal C}_1$ and 
${\cal C}_2$ are equal. Then, if  each irreducible representation of $so_\kappa(3)$ is denoted by  $U_j$ or simply $j$, where $j$ is a real parameter, the irreducible representations of the direct sum 
$so_{\kappa}(3)\oplus so_{\kappa}(3)$ are tensor products
$U^A_j\otimes U^B_j$. Therefore the states of a representation will be
labeled by
\begin{equation}\label{psinot}
\Psi^A_{j,a_3}\otimes \Psi^B_{j,b_3}, 
\end{equation}
where $\Psi^A_{j,a_3}$ ($\Psi^B_{j,b_3}$) belongs to the $U^A_j$ ($U^B_j$) representation such that $a_3, b_3$ are the eigenvalues of $A_3,B_3$, respectively, while $j$ determines de eigenvalue of ${\cal C}_1$ and ${\cal C}_2$, which is the same.

In the space of the functions (\ref{psis}), let us call the functions of the representation $j$ in the form $\Psi^{\kappa,A}_{j,\omega,\ell}$ and 
$\Psi^{\kappa,B}_{j,\omega,\ell}$.
Then
\[
A_3 \Psi^{\kappa,A}_{j,\omega,\ell} = a_3 \Psi^{\kappa,A}_{j,\omega,\ell}=
\frac12(-\omega +\ell \kappa) \Psi^{\kappa,A}_{\omega,\ell}\,,
\qquad
B_3 \Psi^{\kappa,B}_{j,\omega,\ell} = b_3 \Psi^{\kappa,B}_{j,\omega,\ell}=
\frac12(\omega +\ell \kappa) \Psi^{\kappa,B}_{\omega,\ell}
\]
If we use the notation  $\lambda=  \ell \kappa$ and $\omega = n\kappa$, then
\begin{equation}\label{1a3b3}
a_3 = \frac12 (-\omega+ \ell\kappa)= \frac12 \kappa  (-n+\ell),\qquad
b_3 = \frac12 (\omega+  \ell\kappa)= \frac12\kappa(n+\ell)
\end{equation}
where the values of $n$ and $\ell$ will depend on the representation $j$. 
Therefore, the notation of the states in the representation $j$ is
\begin{equation}\label{2a3b3}
\begin{array}{l}
\Psi^{A}_{j,a_3}:= \Psi^{\kappa,A}_{j,\omega,\ell}\,, \qquad a_3 = \frac12 (-\omega+ \ell\kappa)
\\[2ex]
\Psi^{B}_{j,b_3}:= \Psi^{\kappa,B}_{j,\omega,\ell}\,, \qquad b_3 = \frac12 (\omega+  \ell\kappa)
\\[2ex]
\Psi^A_{j,a_3}\otimes \Psi^B_{j,b_3}:= \Psi_{j\otimes j, \omega,\lambda}
\end{array}
\end{equation}
In this notation the action of $A_\pm$ and $B_\pm$ is of the form
\begin{equation}\label{3a3b3}
A_\pm: \Psi_{j\otimes j,\omega,\lambda} \ \to \  
 \Psi_{j\otimes j,\omega\mp \kappa,\lambda\pm \kappa}\,,
 \qquad
 B_\pm: \Psi_{j\otimes j,\omega,\lambda} \ \to \  
 \Psi_{j\otimes j,\omega\pm \kappa,\lambda\pm \kappa}
\end{equation}

\item[(ii)]  

One can check that the action of the Casimir on these states are related to the action of the Hamiltonian in the following way:
\begin{equation}\label{energy0}
\mathcal{H}_{\rm HO}^{\kappa,\omega}\Psi_{j\otimes j,\omega,\lambda}=
\left(\frac12({\cal C}_1+{\cal C}_2) - 
 \frac{\omega^2 -\kappa^2}{\kappa}\right) \Psi_{j\otimes j,\omega,\lambda}
\end{equation}
Therefore,  the eigenvalues of $\mathcal{H}_{\rm HO}^{\kappa}$ are determined by (a) the  frequency $\omega$ of the HO potential, and (b) the eigenvalue of the Casimir for the representation of  $so_\kappa(4)$. We can say that the representation of 
$so_\kappa(4)$ is the direct sum of subspaces with different frequency. Each eigenspace is selected by choosing this frequency.

\item[(iii)]  {Symmetries in terms of spherical shift operators}

We have characterized the shift operators and the space where they act.
The Hamiltonian ${\mathcal{H}}_{\rm HO}^{\kappa}$ (\ref{haes}) given in terms
of $a^\pm_i$ can be expressed in terms of $A_\pm$ and $B_\pm$
as shown above (\ref{energy0}). The symmetries are the operators of $so_\kappa(4)$ that keep invariant the frequency $\omega$. The simplest of them are 
\[
L_3\,,\quad
\tilde Q_{AA}= A_+ A_-\,,\quad  \tilde Q_{BB}= B _+ B_-\,,\quad
\tilde Q_{AB}= A_+ B_+ 
\]
The three symmetries $\tilde Q_{AB}, \tilde Q_{AB}, \tilde Q_{AB}$ are independent, which shows that our two-dimensional system is indeed superintegrable. They are equivalent to the FD tensor components in polar coordinates.

\end{itemize}

The space of states (the  support space), the eigenfunctions and eigenvalues, will depend on the representation, thus, we have to specify each case $\kappa>0$, $\kappa<0$ and $\kappa=0$ by separate.
We have not a general form of representations valid for any $\kappa$.

\subsection{The case $\kappa>0$. Eigenvalues and degeneracy}


From (\ref{2a3b3}) and(\ref{3a3b3}), the action of $A_\pm$ and $B_\pm$ on the eigenfunctions $\Psi_{j\otimes j,n ,\ell}:=
\Psi_{j\otimes j,n\kappa,\ell\kappa}$
is
\begin{equation}\label{abpm0}
A_\pm \Psi_{j\otimes j,n ,\ell}\propto \Psi_{j\otimes j,n\mp 1 ,\ell\pm1}\,,\qquad
B_\pm \Psi_{j\otimes j,n ,\ell}\propto \Psi_{j\otimes j,n\pm1 ,\ell\pm1}
\end{equation}

\begin{figure}[h]
	\centering
\includegraphics[width= 6 cm]{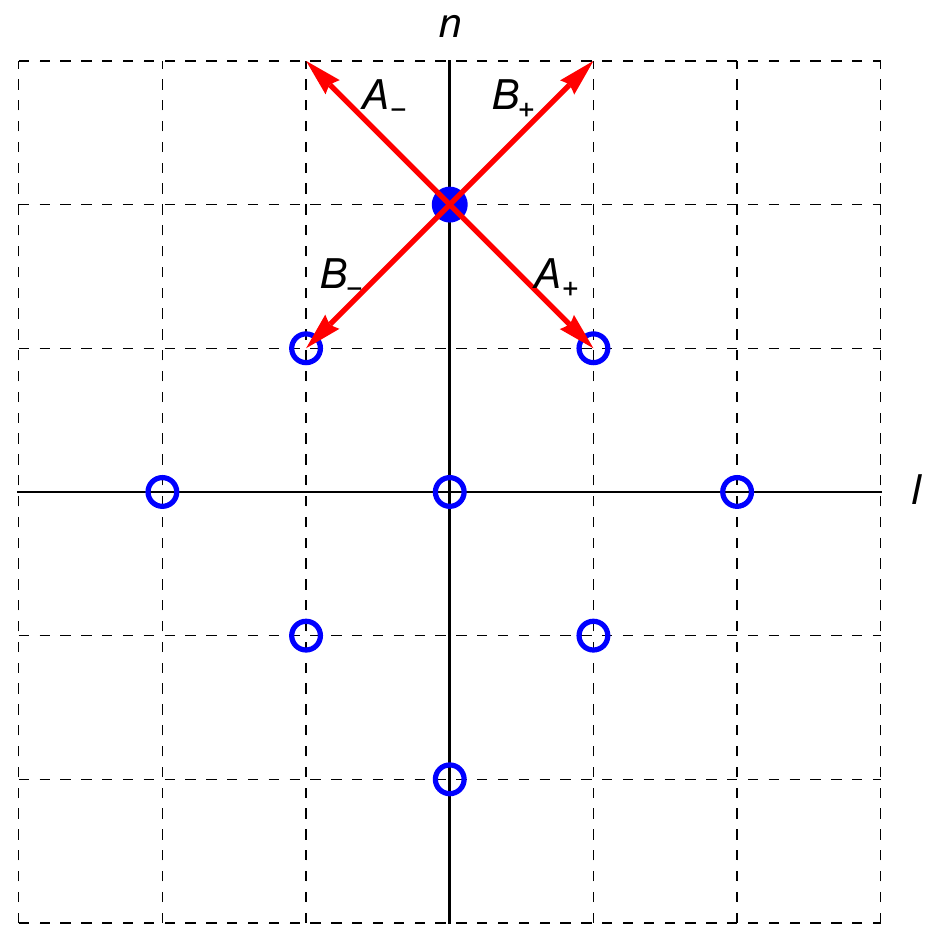}
\caption{\small  The points represent the states of a presentation of $so(4)$: $j\otimes j$, with $j=1$ (or $\bar n=2$) in this case. The action of the operators $A_\pm$ and $B_\pm$ is represented by arrows. The fundamental state is in blue.
\label{f1a}}
\end{figure}

From the commutation rules (\ref{sop1})-(\ref{sop2}), the generators $\langle A^\pm, A_3\rangle\oplus \langle B^\pm,  B_3 \rangle$ close a $so(3)\oplus so(3)\approx so(4)$ Lie algebra. In this case $j$ designs the irreducible representations of $so(3)$, therefore, $j$ will be a half-integer positive number. 
In this case we will take the fundamental state $\Psi_{j\otimes j,\bar n ,0}$ annihilated by $A_-$ and $B_+$. Since the operators $A_- $ and $B_+$ increase the frequency $\omega$, this fundamental state is characterized by a highest frequency $\omega_{\max}= \bar n \kappa$ and $\ell=0$, state of the representation $j=\bar n/2$
for each $so(3)$ subalgebra. We will use the following notation and properties for this highest weight:
\[
\begin{array}{c}
{\rm highest\ weight:}\ 
\Psi_{j\otimes j,\bar n ,0} = N_0\,\sin^{1/2} \theta \cos^{\bar{n}+1/2} \theta\,,
\qquad \omega_{\max}= \bar n \kappa\,,\  
\quad \ell= 0\,,\ 
j= \bar n/2\,,
\\[2.5ex]
A_- \Psi_{j\otimes j,\bar n,0} = B_+ \Psi_{j\otimes j,\bar n,0} =0
\end{array}
\]
The dimension of each representation $j\otimes j$ is $(\bar n +1)^2=(2j+1)^2$, $\bar n=0,1,2,\dots$.
The basis is given by the wavefunctions obtained by the action of $A_+, B_-$ which decrease the values of $\omega$, as shown in (\ref{abpm0}): 
\[
(A_+)^q(B_-)^p\Psi_{j,\bar n,0}\propto \Psi_{j,n ,\ell}\,,\quad (n ,\ell) = (\bar n,0) + p(-1, -1)+q(-1,1),\quad p,q=0,1,2\dots\ |n|,|\ell|\leq \bar n
\]
Inside the representation $j\otimes j$ of this algebra $so(4)$, the range of frequencies run from $\bar \omega = \bar n \kappa$ up to ${\omega} = - \bar n \kappa$, see Fig.\ref{f1a}.
One conclusion is that the frequencies must be  multiples of the curvature:
$\omega = n \kappa$, otherwise the operators $A_\pm,B_\pm$ would not correspond to a unitary representation. 

\begin{figure}[h]
	\centering
\includegraphics[width= 4.5 cm]{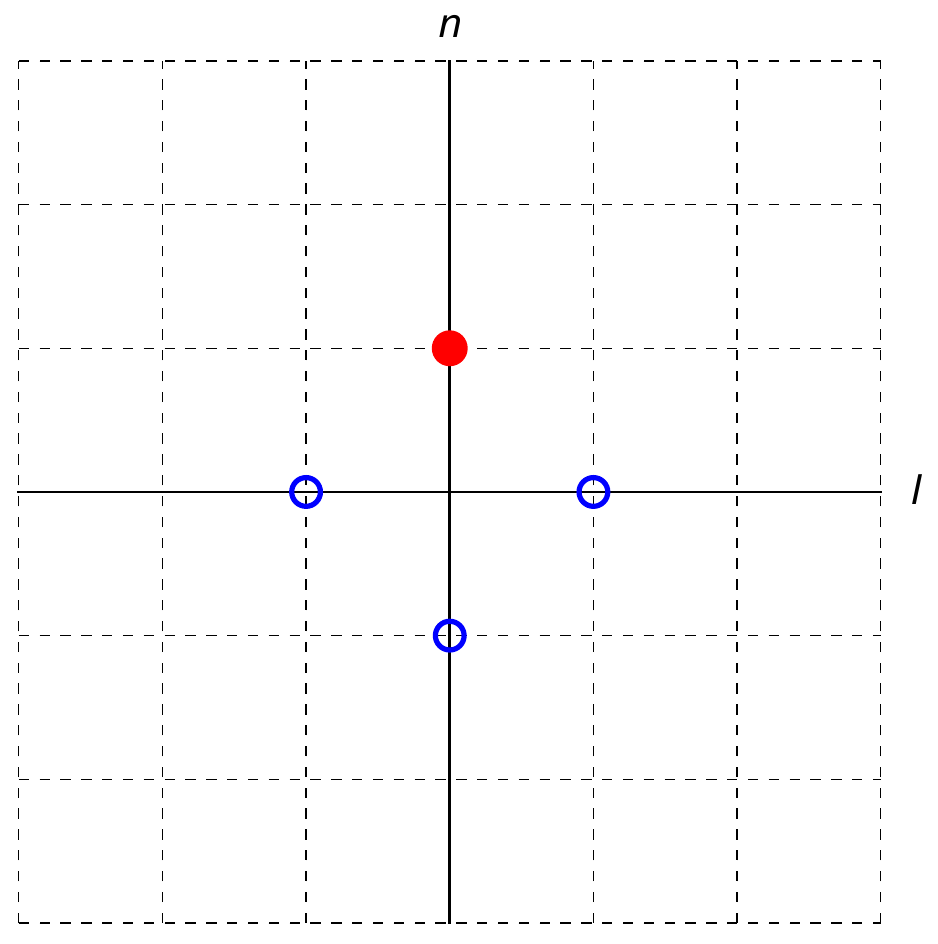}
\includegraphics[width= 4.5 cm]{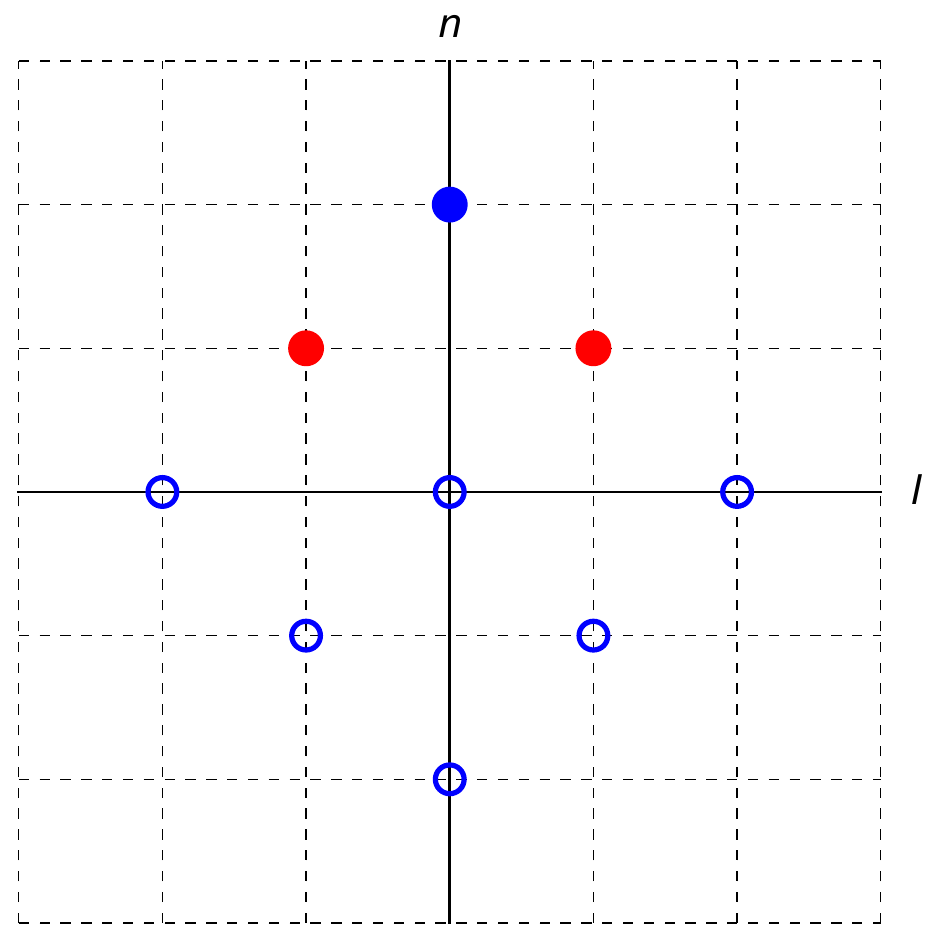}
\includegraphics[width= 4.5 cm]{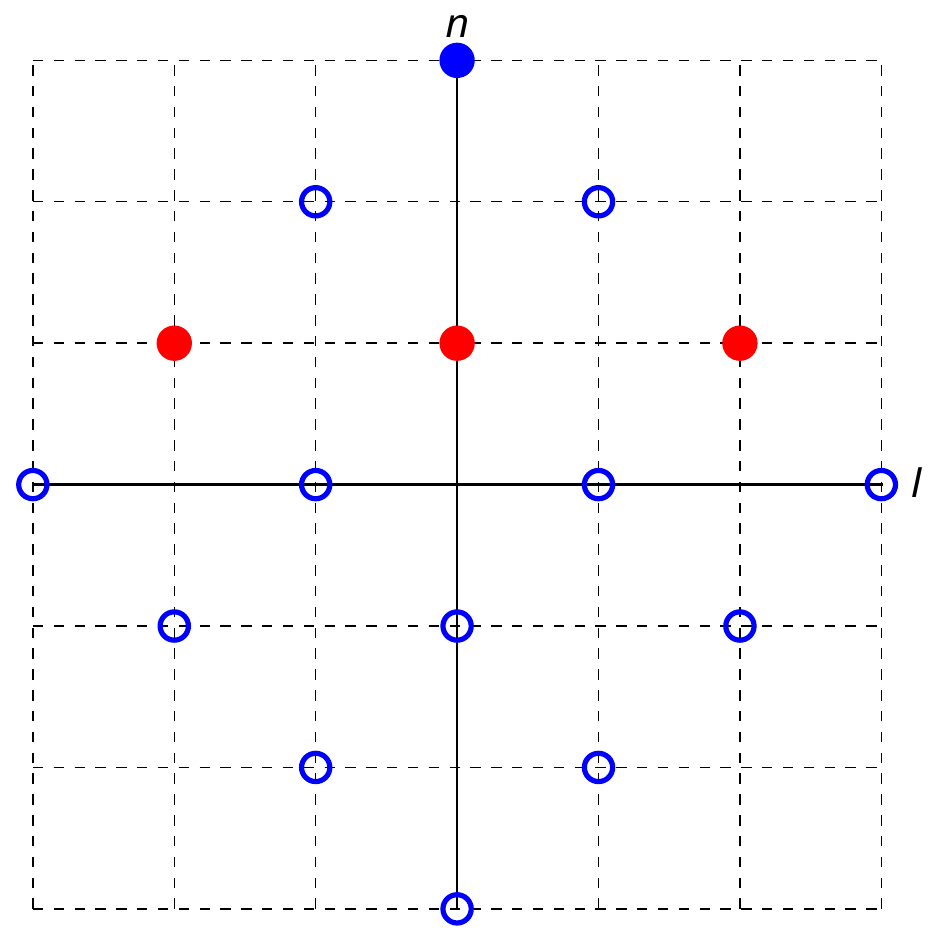}
\caption{\small Each graphic represents the eigenstates of an energy level of the HO for $\kappa>0$ and frequency $\omega_0=\kappa$. The left is for the ground energy with one state (in red), the center is for the first excited level (with two states in red) and the right for the second level (with three states in red). Each energy level corresponds to a different $so(4)$ representation: $j=1/2$, $j=1$ and $j=3/2$ (or $\bar n=1$, $\bar n=2$, $\bar n=3$), respectively.
\label{f1b}}
\end{figure}

\begin{figure}[h]
	\centering
\includegraphics[width= 8 cm]{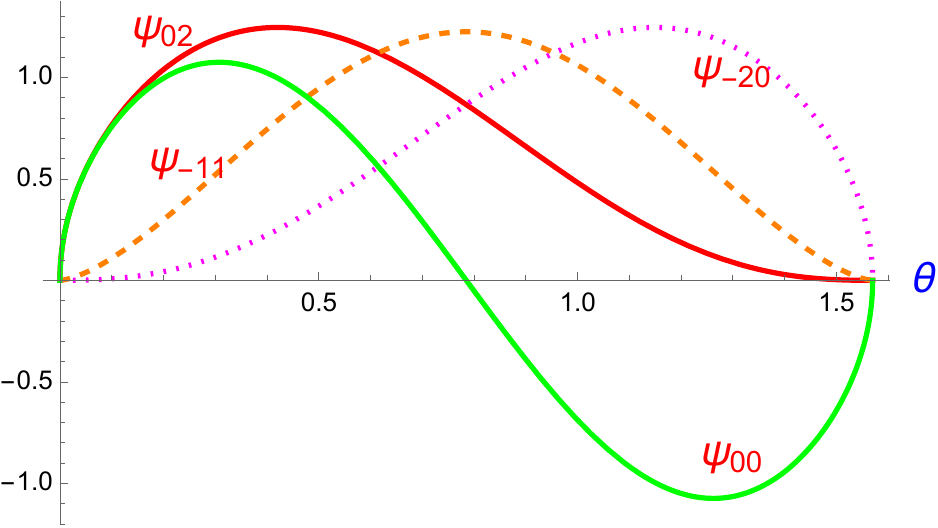}
\caption{\small  Plot of the functions $\Psi_{j=1,n=2,l=0}$ (blue),
$\Psi_{j=1,n=1,l=-1}$ (dashed), $\Psi_{j=1,n=0,l=-2}$,  $\Psi_{j=1,n=0,l=0}$ (green).
\label{f1c}}
\end{figure}

Consider a particular frequency $\omega_0 = n_0 \kappa$. 
This frequency will belong to a representation $j = \bar n/2$, as far as 
$\bar n\geq n_0$, or $\bar n = n_0 + n$.
Then, we can answer to the fundamental question: Fixed any positive 
frequency $\omega_0 = n_0 \kappa$, what are the eigenvalues (energies) and degeneracy of the eigenstates of the Hamiltonian with such frequency?
\begin{itemize}
\item
The representations which include the frequency $\omega_0$ are given
by $j=\bar n/2$, such that $\bar n = n_0+n$, $n=0,1,2,\dots$  See Fig.~\ref{f1b}
\item
The values of the energy, corresponding to the $n$-excitation level, belong  to the representation $\bar n = n_0+n$. According to (\ref{energy0}) they are given by (see also \cite{bonatsos94})
\[
E^n_{\omega_0,\kappa}= 
\frac4{\kappa}\left(\frac{\omega_0+ n\kappa}2 \Big(\frac{\omega_0+ n\kappa}2 +\kappa\Big)\right) - \frac{\omega_0^2 -\kappa^2}{\kappa} 
= 2\omega_0(n+1) + \kappa (n+1)^2 \,, \quad n=0,1,2,\dots
\]
\item
The degeneracy of the above eigenvalue is $D=n+1$, as can be appreciated from the graphic of Fig.~\ref{f1b}.

\end{itemize}
Then, the energy of a state $\Psi_{j_0\otimes j_0,n_0, \ell}$, depends on the representation $j_0\otimes j_0$ to which it belongs. 
Let $j_0= n_1/2$ and  $n_1- n_0 =n$; therefore the energy will be
$E^{n_1-n_0}_{\omega_0,\kappa}$. In summary, the energy levels for the spherical oscillator of frequency $\omega_0$ are associated to  representations of the algebra $so(4)$ algebra. Each of these representations is characterized by a frequency 
$\omega=\bar n \kappa$ which must be greater than $\omega_0$, i.e. $\omega-\omega_0\geq 0$. Some eigenfunctions are plotted in Fig.~\ref{f1c}.

\subsection{The case $\kappa<0$. Eigenvalues and degeneracy}

In this case, the generators $\langle A_\pm, A_3\rangle\oplus \langle B_\pm,  B_3 \rangle$ of $so_\kappa(4)$, $\kappa<0$, close a Lie algebra isomorphic to $so(2,2)\approx so(2,1)\oplus so(2,1)$ . Its relevant representations here are also of the form $j\otimes j$, where
$j$ is a  positive real number and $U_j$ represents a discrete $so(2,1)$ representation.
The action of $A_\pm$ and $B_\pm$ on an eigenfunction $\Psi_{j,n ,\ell}$
according to (\ref{1a3b3}) and (\ref{2a3b3}) is
\begin{equation}\label{abpm}
A_\pm \Psi^A_{j,n ,\ell}\propto \Psi^A_{j,n\pm 1 ,\ell\mp1}\,,\qquad
B_\pm \Psi^B_{j,n ,\ell}\propto \Psi^B_{j,n\mp1 ,\ell\mp1}
\end{equation}

\begin{figure}[h]
	\centering
\includegraphics[width= 6 cm]{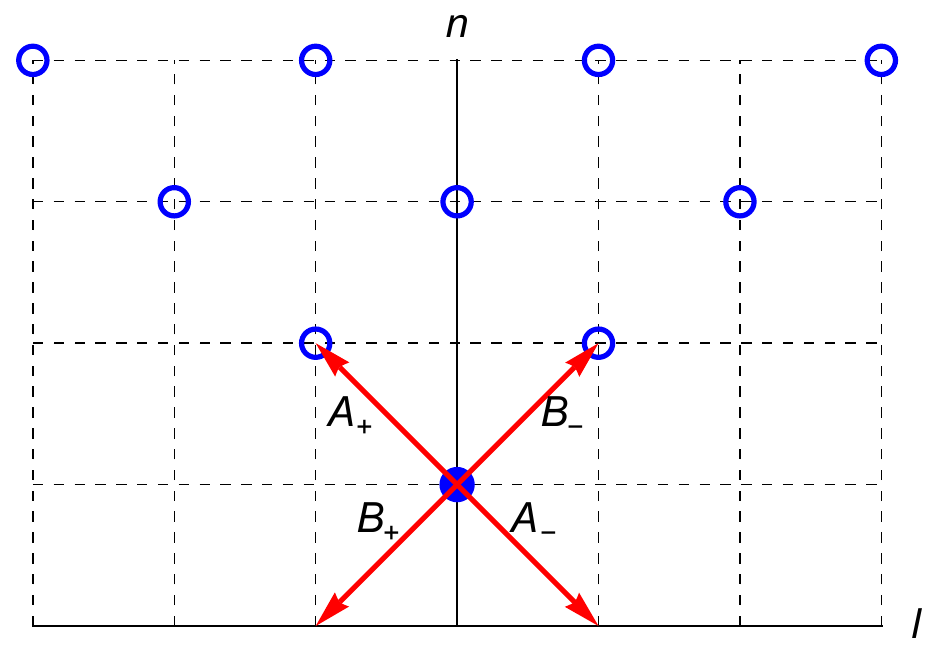}
\caption{\small  The points represent the states of a presentation of $so(2,2)$: $j\otimes j$, with $j=1/2$ (or $\bar n=1$) in this case. The action of the operators $A_\pm$ and $B_\pm$ is represented by arrows. The fundamental state is in blue.
\label{f2a}}
\end{figure}

In this case, let us call the positive fundamental state, characterized by $n=\underline{n}$ and $\ell = 0$, as follows
\[
\begin{array}{c}
\Psi_{j\otimes j,\underline{n} ,0} = 
N_0\,\sinh^{1/2} \theta \cosh^{-\underline{n}-1/2} \theta,
\quad \omega_{\rm min}= \underline{n} |\kappa|\,,\quad  j= \underline{n}/2\,,
\\[2.5ex]
A_- \Psi_{j,\underline{n},0} = B_+ \Psi_{j,\underline{n},0} =0\,,\quad 
\underline{n}>0,\ j=\underline{n}/2>0
\end{array}
\]
Since now $A_-$ and $B_+$ are decreasing frequency operators, $\omega_{\rm min}$ is a minimum frequency.
The basis of the representation is given  by the wavefunctions, obtained by the action of $A_+, B_-$, which increase the values of $\omega$, as shown in (\ref{3a3b3}): 
\[
(A_+)^q(B_-)^p\Psi_{j\otimes j,\underline{n} ,0}\propto \Psi_{j\otimes j,n ,\ell}\,,\quad (n ,\ell) = (\bar n,0) + p(1, 1)+q(1,-1),\quad p,q=0,1,2\dots
\]
with $p+q=n-\underline{n}$, see Fig.~\ref{f2a}.
 
%

\begin{figure}[h]
	\centering
\includegraphics[width= 5 cm]{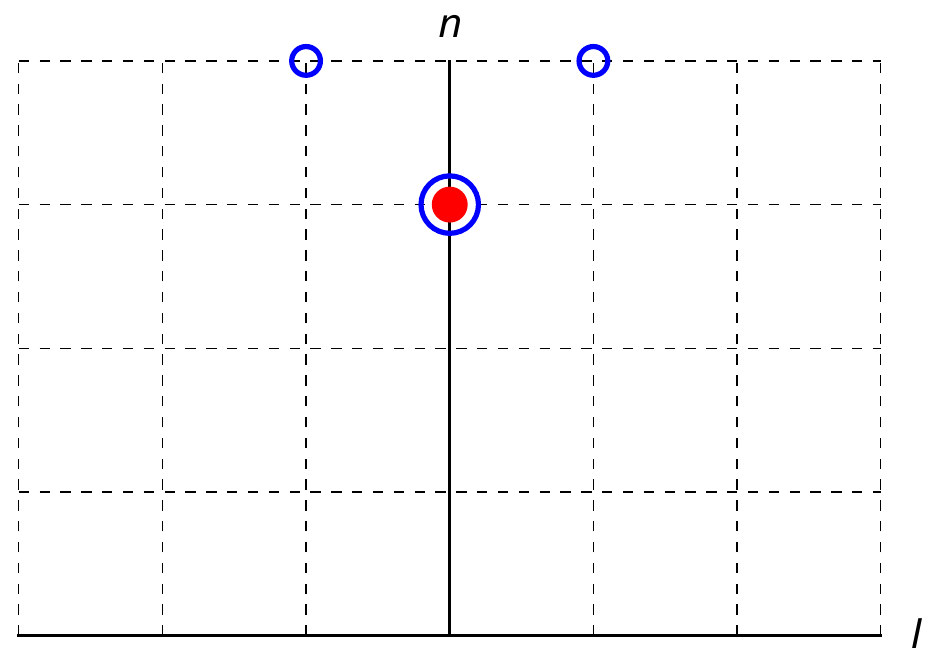}
\includegraphics[width= 5 cm]{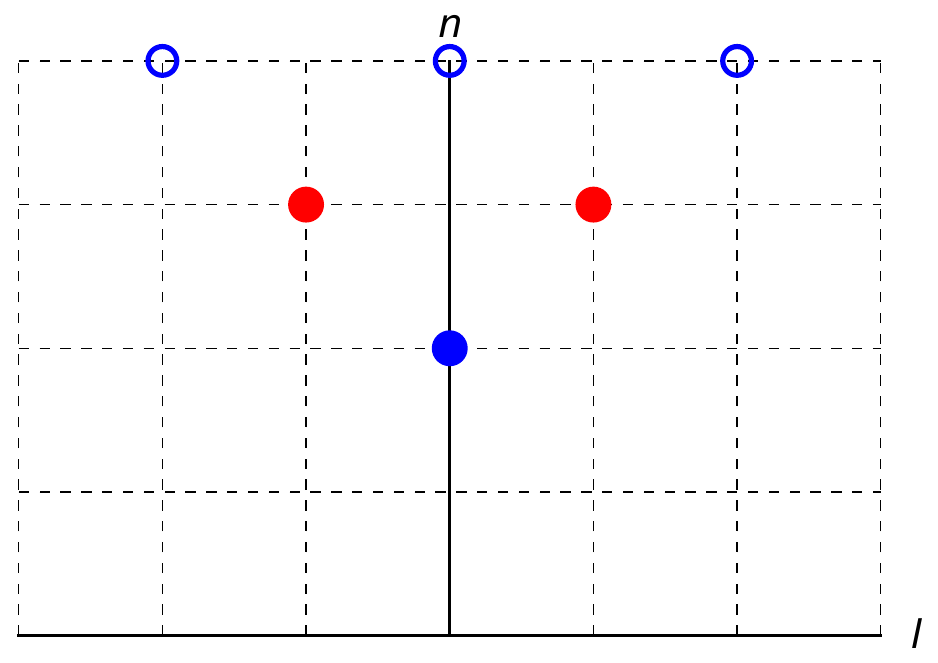}
\includegraphics[width= 5 cm]{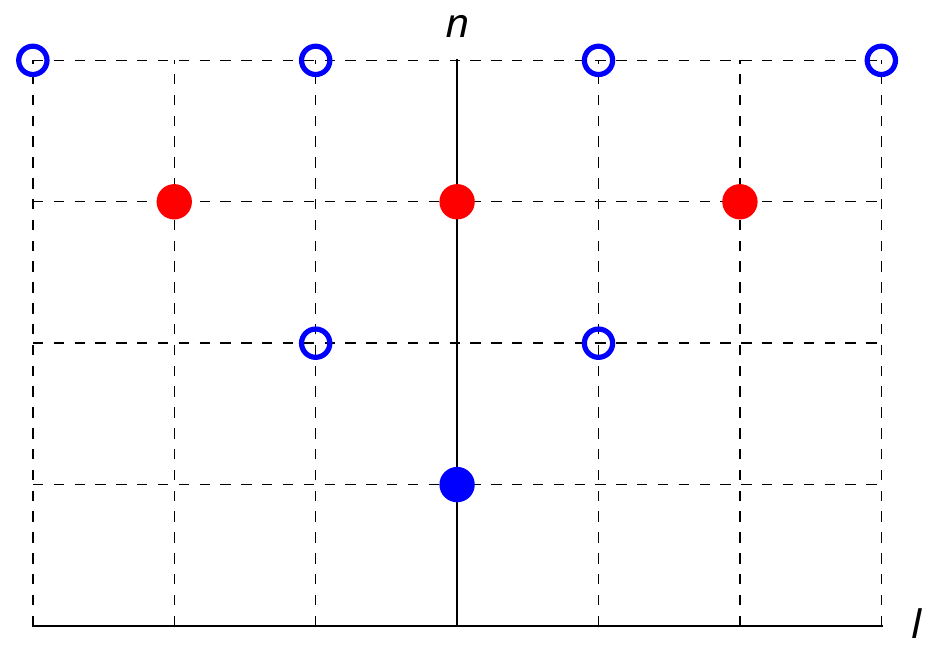}
\caption{\small Each graphic represents the eigenstates of an energy level of the HO for $\kappa<0$ and frequency $\omega_0=3\kappa$. The left is for the ground energy with one state (in red), the center is for the first excited level (with two states in red) and the right for the second level (with three states in red). Each energy level corresponds to a different $so(4)$ representation: $j=3/2$, $j=1$ and $j=1/2$ (or ${\underline n}=3$, ${\underline n}=2$, ${\underline n}=1$), respectively.
\label{f2b}
}
\end{figure}

Consider a particular positive frequency $\omega_0 = n_0 |\kappa|$. 
This frequency will belong to the representation $j = \underline{n}/2$, as far as 
$|\underline{n}|\leq |n_0|$, or $n_0 =  \underline{n} + p$, where $p$ is a positive integer. Therefore, the number of representations including $\omega_0$ will be finite, see Fig.~\ref{f2b}:
\[
j_0= n_0/2,\  j_1= (n_0 - 1)/2,\  \dots \ 
j_p= (n_0 - p)/2,\dots\  j_{\rm min}=  (n_0 - p_{\rm max})/2,\quad
{\rm where}\quad 0< n_0- p_{\rm max}\leq  1
\]
or in terms of frequencies
\[
\omega_0=n_0|\kappa|,\  \omega_1= (\omega_0 +\kappa),\  \dots \ 
\omega_p= (\omega_0 +p\kappa),\dots\  \omega_{\rm min}=  (\omega_0 + p_{\rm max}\kappa),\quad
{\rm where}\quad 0< n_0- p_{\rm max} \leq  1
\]

\begin{figure}[h]
	\centering
\includegraphics[width= 8 cm]{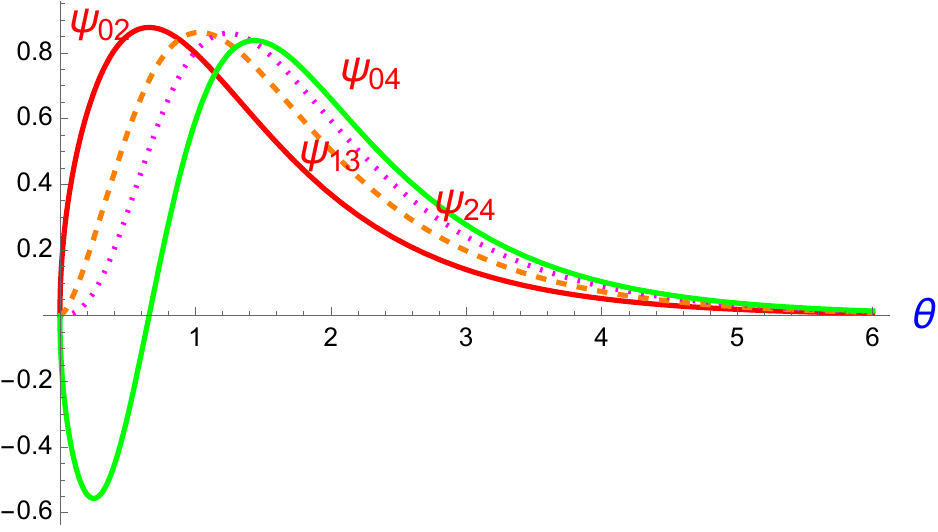}
\caption{\small Plot of the functions for $\kappa<0$: $\Psi_{j=1,n=2,l=0}$ (blue),
$\Psi_{j=1,n=3,l=1}$ (dashed), $\Psi_{j=1,n=4,l=2}$ (dotted) and $\Psi_{j=1,n=4,l=0}$ (green). They correspond to some of the points of graphic in the centre of Fig.~\ref{f2b}.
\label{f2c}}
\end{figure}

Therefore, fixed any positive 
frequency $\omega_0=n_0|\kappa|$, the energies and degeneracy of the states with such frequency are as follows. The values of the energy, corresponding to the $p$-level, belong  to the representation $j_p = (n_0-p)/2$. According to (\ref{energy0}) they are given by
\[
E^p_{\omega_0,\kappa}= 
\frac4{\kappa}\left(\frac{\omega_0+ p\kappa}2 \Big(\frac{\omega_0+ p\kappa}2 +\kappa\Big)\right) - \frac{\omega_0^2 -\kappa^2}{\kappa} 
= 2\omega_0(p+1) + \kappa (p+1)^2 \,, \quad p=0,1,2,\dots p_{\rm max}
\]
The degeneracy of the above eigenvalue is $D=2 p+1$.
Then, the energy of a state $\Psi_{j,n_0, \ell}$, depends on the representation to which it belongs. The representation of $\Psi_{j\otimes j,n_0, \ell}$ is $j$, then
$j=\bar n/2$ and  $2j- n_0 =n$; therefore the energy will be
$E^{2j-n_0}_{\omega_0,\kappa}$. Some examples of eigenfunctions are shown in Fig.~\ref{f2c}.

\subsection{The case $\kappa=0$. Eigenvalues and degeneracy}

When we perform the limit $\kappa\to 0$, we obtain the following differential realization 
\begin{equation}\label{abs0}
\begin{array}{l}
\displaystyle
A_+= \frac12  \left(-\partial_\theta 
-\frac{1/2+i \partial_\phi}{\theta} + \omega  \theta \right)e^{i\phi} 
\\[2.ex]
\displaystyle
A_-= \frac12 e^{-i\phi}  \left(\partial_\theta 
-\frac{1/2+i \partial_\phi}{ \theta} + \omega  \theta \right)
\\[2.ex]
\displaystyle
A_3 =-\frac{\omega}2   
\\[2ex]
\displaystyle
B_-= \frac12 e^{-i\phi} \left(-\partial_\theta 
+\frac{1/2+i \partial_\phi}{ \theta} +  \omega  \theta \right) 
\\[2.ex]
\displaystyle
B_+= \frac12  \left(\partial_\theta 
+\frac{1/2+i\partial_\phi}{ \theta} + \omega  \theta \right) e^{i\phi}
\\[2.ex]
\displaystyle
B_3=\frac{\omega}2   
\end{array}
\end{equation}

These are exactly the flat formulas (\ref{ABs11}) and (\ref{ABs22}) of section 2, where instead of $r$ now we have $\theta$. The commutation relations become
\begin{equation}\label{sop10}
\displaystyle
[A_-,A_+]=  {\omega},
\quad
[A_3,A_\pm]= 0,
\end{equation}
\begin{equation}\label{sop20}
\displaystyle
[B_-,B_+]=  -{\omega},
\quad
[B_3,B_{\pm}]= 0 
\end{equation}
They constitute the direct sum of two Heisenberg algebras, where the generator
$iJ_3=L_3$ acts in a semidirect way (\ref{sop3}). The commuting symmetries become
\[
\{\ \omega \,, L_3\,, \mathcal{H}_{\rm HO}^{\kappa} \}
\] 
where the frequency now is constant and the states belong to the same representation
 determined by the fundamental state.  
In this case the eigenstates are characterized by the energy $E^n_\omega$ and the angular momentum $\ell$,
\[
\mathcal{H}_{\rm HO}^{0}\Psi_{\omega_0,n,\ell}= 
E^n_\omega\,\Psi_{\omega,n,\ell}\,,\qquad
L_3\Psi_{\omega,n,\ell}= 
\ell\,\Psi_{\omega,n,\ell}
\]

The Hamiltonian is given by
 (\ref{hhh2b}) in terms of the operators $a_i^\pm$. 
 This can also be expressed in terms of $A_\pm$ and $B_\pm$ as follows
 \begin{equation}\label{hhh2c}
 \begin{array}{l}
 \displaystyle
\mathcal{H}_{\rm HO}^{\kappa} = 4A_+ A_-  \,
+\frac{4}{\kappa}(A_3(A_3-\kappa)  -\frac{\omega^2-\kappa^2}{\kappa}
\\[2.ex]
\displaystyle
\qquad = 4A_+ A_- - 2i \omega J_3
+(2\omega +\kappa) -\kappa J_3^2
\end{array}
\end{equation} 
If we take the limit $\kappa \to 0$ of this expression we get
\begin{equation}\label{hhh2d}
\mathcal{H}_{\rm HO}^{0} =  
4A_+ A_- - 2i \omega J_3
+ 2\omega 
\end{equation} 
which can also written as follows
\[
\mathcal{H}_{\rm HO}^{0}\Psi_{j,\omega,\lambda}=
2\, (A_-(\omega)A_+(\omega)+
B_-(\omega)B_+(\omega ))
\]

The ground state is characterized by $\omega$ and $\ell=0$,
\[
\Psi_{\omega,n=0,\ell=0} = N_0\, e^{-\omega^2 \theta/2}\sqrt{\theta}\,,\qquad L_3  \Psi_{\omega,0,0}=0\,,\qquad
A_- \Psi_{\omega,0,0} = B_+ \Psi_{\omega,0,0} = 0
\]

\begin{figure}[h]
	\centering
\includegraphics[width= 6 cm]{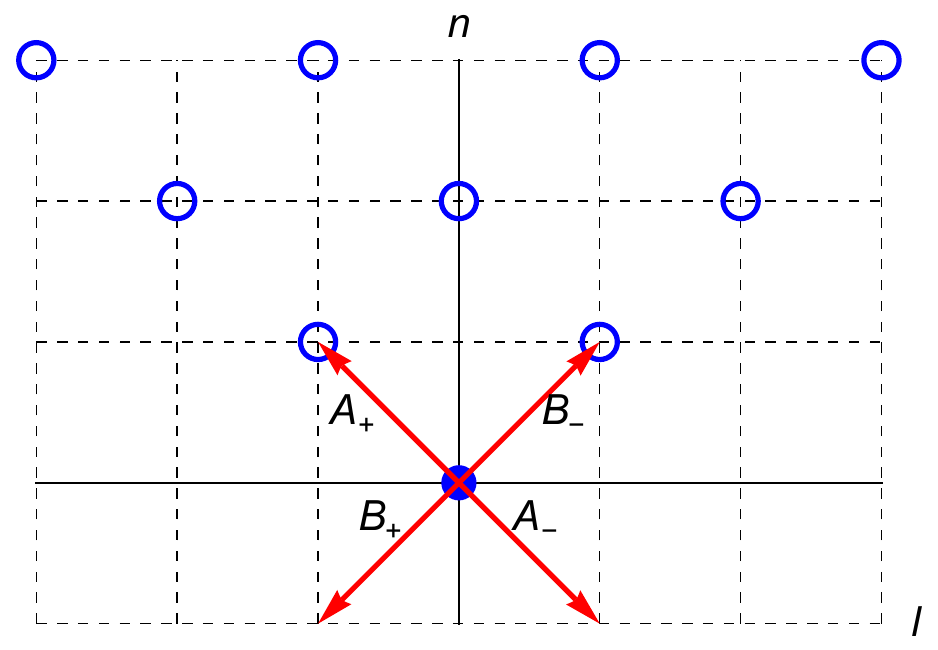}
\includegraphics[width= 6 cm]{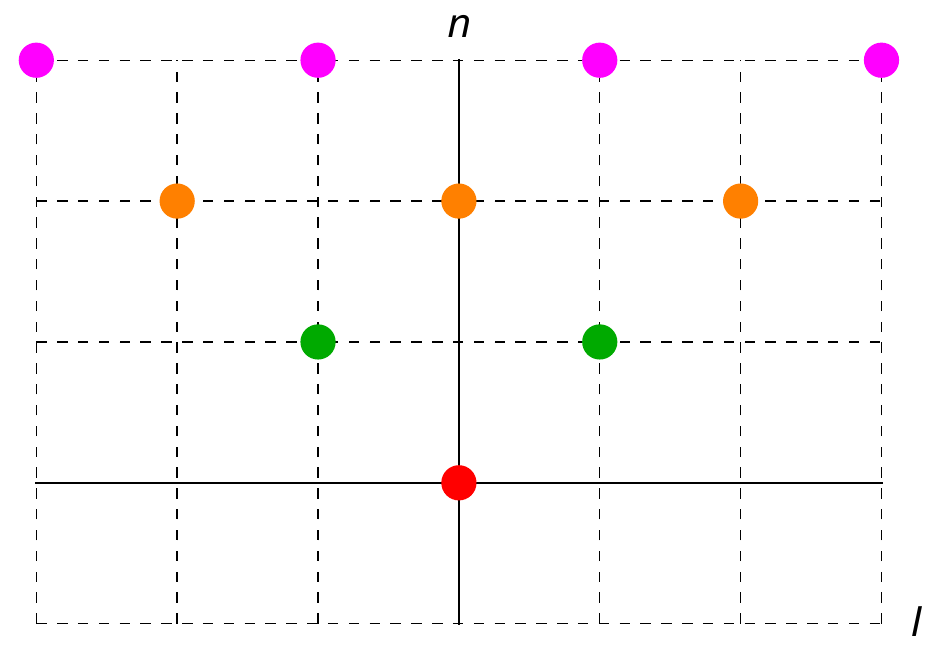}
\caption{\small (Left) The points represent the states of the presentation of $so_0(4)$ with $\omega_0$ and $l_0 = 0$ . The action of the operators $A_\pm$ and $B_\pm$ is represented by arrows. The fundamental state is in blue. (Right) The states of energy levels are represented by point with the same color. All the states belong to the same representation.
\label{f7}
}
\end{figure}

The rest of vectors of the basis for the support space of the direct sum representation are constructed as follows, see Fig.~\ref{f7},
\[
(A_+)^q(B_-)^p\Psi_{\omega,0,0}\propto \Psi_{\omega,n,\ell}\,,\quad (n ,\ell) =  p(1, 1)+q(1,-1),\quad p,q=0,1,2\dots
\]
The eigenvalues and eigenvectors are obtained by applying $A_+$ and $B_-$,
as usual. The formula this time is
\[
E^n_{\omega,\ell}  
= 2\omega(n+1)   \,, \qquad n=0,1,2,\dots
\]
where $n$ is the level of energy. The degeneracy of each level
is $n+1$.

\begin{figure}[h]
	\centering
\includegraphics[width= 8 cm]{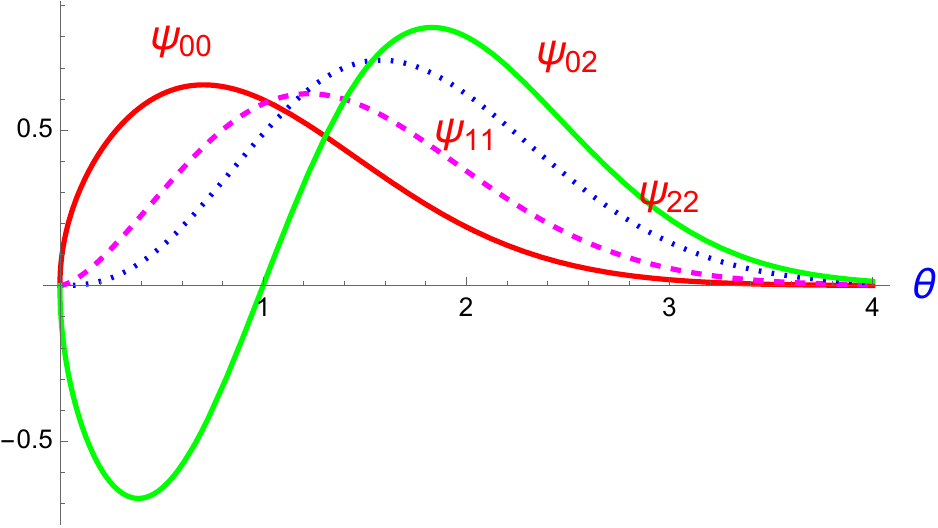}
\caption{\small Plot of the functions for $\kappa=0$: $\Psi_{n=0,l=0}$ (Red),
$\Psi_{n=1,l=1}$ (dashed), $\Psi_{n=2,l=2}$ (dotted) and $\Psi_{n=2,l=0}$ (green). They correspond to some of the points of graphic in the right of Fig.~\ref{f7}.
\label{f8}}
\end{figure}

\sect{Symmetries of the classical curved HO}

{\it Classical Hamiltonian and symmetries}
\medskip

In the classical frame, the curved HO is defined on the same surface $\Sigma_\kappa$
given in (\ref{ksphere})
characterized  by the curvature parameter $\kappa$. The classical fields are defined by
\begin{equation}\label{js}
{\rm J}_{0i} = x_0 p_i- \kappa\, x_i p_0,\qquad
{\rm J}_{jk} = x_j p_k -  x_k p_j,\qquad {\rm J}_{jk} = -{\rm J}_{kj} \qquad j\neq k;\  i,j,k>0
\end{equation}
where $x_k$ and $p_k$ are conjugate canonical variables $\{x_i,p_j\} =\delta_{ij}$. These generators fulfil  the Poisson Brackets (PB), 
\begin{equation}\label{comjsC}
\{ {\rm J}_{0i},{\rm J}_{ik} \} =  -{\rm J}_{0k}\,\quad 
\{ {\rm J}_{ij},{\rm J}_{jk} \} = -{\rm J}_{ik}\,,\quad
\{ {\rm J}_{0i},{\rm J}_{0j} \} = \kappa {\rm J}_{ij}
\end{equation}
The kinetic term is the free Casimir function
\begin{equation}\label{cas}
{\rm H}^\kappa_0={\rm C}_\kappa =  \sum_{i} {\rm J}_{0i}^2 + \kappa \sum_{i<j} {\rm J}_{ij}^2 \,, \qquad i,j=1,2,3
\end{equation}
while the potential has the form
\begin{equation}\label{pots}
 {\rm V}^\kappa_{\rm HO}= {\omega^2
} \,
\frac{ {\bf x}^2 }{ {x_0}^2 }= 
{\omega^2
} \,
\frac{ {\bf x}^2 }{ 1-\kappa {\bf x}^2 } 
\end{equation}
Then, the classical Hamiltonian is
\begin{equation}\label{hhh2C}
\mathcal{H}_{\rm HO}^{\kappa} =  \sum_{i} {\rm J}_{0i}^2 
+ {\omega^2} \,
\frac{ {\bf x}^2 }{ {x_0}^2 }
+ \kappa \sum_{i<j} {\rm J}_{ij}^2 \,,
\end{equation} 
The basic symmetries of this Hamiltonian are

\begin{equation}\label{i0k}
{\rm I}_{ij}= {\rm J}_{ij}^2\,,\qquad 
{\rm I}_{0i}^{\kappa}= {\rm J}_{0i}^2 
+{\omega^2}{ \mathrm{T}_{\kappa}^2(\varphi_{0i})} 
\end{equation}
The factorization functions in terms of ambient coordinates are
given by
\begin{equation}\label{aasC}
{\rm a}^\mp_{0i}(\omega)=\pm i\, {\rm J}_{0i} +{\omega}\,\mathrm{T}_{\kappa}(\varphi_{0i}),\qquad
 \mathrm{T}_{\kappa}(\varphi_{0i})=\frac{x_i}{x_0}
\end{equation}
It is clear that the symmetry function is factorized in the form
\[
{\rm I}_{0i}^{\kappa}={\rm a}^+_{0i}(\omega) {\rm a}^-_{0i}(\omega)
\]
The general symmetries take the form 
\begin{equation}\label{sym4jcx}
{\rm Q}_{ij}^{\kappa}:={\rm a}^+_{0i}(\kappa) {\rm a}^-_{0j}(\kappa),\quad i,j=1,2
\end{equation}

However, we will adopt the following point of view. In order to take into account
$\omega$ we consider a new variable $\zeta$ and its conjugate variable will be identified with $\omega$, $p_\zeta:= \omega$. Then, we propose the new factorization
\begin{equation}\label{zetas}
{\rm a}^\mp_{0i}(\kappa,x_0,p_0,x_i,p_i,\zeta,p_\zeta):=
 \left(\pm i\, {\rm J}_{0i} +{p_\zeta}\,\mathrm{T}_{\kappa}(\varphi_{0i})
\right) e^{\pm i \kappa \zeta}
\end{equation}

These factor functions satisfy the following PBs
\begin{equation}\label{aes}
\begin{array}{l} 
\displaystyle
\{{\rm a}^-_{i}(\kappa),{\rm a}^+_{i}(\kappa)\}= - 2 i p_\zeta 
\\[2.ex]
\displaystyle
\{{\rm a}^-_{i}(\kappa),{\rm a}^+_{j}(\kappa)\}=
\{{\rm a}^+_{i}(\kappa),{\rm a}^-_{j}(\kappa)\}= 2 \kappa {\rm J}_{ij}
\\[2.ex]
\displaystyle
\{p_\zeta\,,{\rm a}^\pm_{i}(\kappa)\}= \pm i \kappa {\rm a}^\pm_{i}(\kappa)
\\[2.ex]
\displaystyle
\{{\rm a}^+_{i}(\kappa),{\rm a}^+_{j}(\kappa)\}=\{{\rm a}^-_{i}(\kappa),{\rm a}^-_{j}(\kappa)\}=0
\\[2.ex]
\displaystyle
\{{\rm J}_{ij},{\rm a}^\pm_{j}(\kappa)\}=
 - {\rm a}^\pm_{i}(\kappa)
\end{array}
\end{equation}
where ${\rm a}^\pm_{0j}= {\rm a}^\pm_{j}$.
These symmetries, as well as the Hamiltonian,  do not depend on $\zeta$, thus the conjugate momentum $p_\zeta=\omega$ is constant in the physical motion.

Since the functions $a^\pm_k$ are complex, we may take the symmetric and antisymmetric combinations in order to get real functions,
\begin{equation}
{\rm F}_{ij}^{\kappa}:= \frac12 \Big({\rm a}^+_{ j}(\kappa,\omega) {\rm a}^-_{ i}(\kappa,\omega) + {\rm a}^+_{ i}(\kappa,\omega) {\rm a}^-_{ j}(\kappa,\omega)\Big),\quad
{\rm D}_{jk}^{\kappa}:= \frac1{2 i} \Big({\rm a}^+_{ j}(\kappa,\omega) {\rm a}^-_{ k}(\kappa,\omega) - {\rm a}^+_{ k}(\kappa,\omega) {\rm a}^-_{j}(\kappa,\omega)\Big)
\end{equation}
Explicitly, they can be written as
\begin{equation}\label{aij}
{\rm F}_{ij}^{\kappa }=
\frac{ x_i x_j (\omega^2-\kappa^2/4)} {x_0^2} 
    +\frac12\,({\rm J}_{0i}{\rm J}_{0j}+{\rm J}_{0j}J_{0i})\,, \qquad 
{\rm D}_{ij}^{\kappa}= -\omega  \, {\rm J}_{ij}
      \end{equation}
The components ${\rm F}_{ij}^{\kappa}$ form the  Demkov--Fradkin tensor  of symmetries for the curved HO.
In the limit $\kappa \to 0$, $x_0 \to 1$, we get the flat Demkov--Fradkin tensor \cite{fradkin65}:  
   \begin{equation}
   {\rm F}^{\kappa=0}_{ij}= {\omega^2 x_i x_j} -  p_{i}p_{j}
   \end{equation}
The DF tensor components are symmetries
 \begin{equation}
  [ {\rm F}_{ij}^{\kappa},{\mathcal{H}}_{\rm HO}^{\omega}]=0
   \end{equation}
\medskip
   
\noindent
{\it Factorization functions in spherical basis}
\medskip
   
   Next, we will write the shift operators in another basis (in two dimensions where $i,j=1,2$) to simplify the algebraic structure as in the quantum case. Let us define it as follows 
\begin{equation}\label{aijps}
\begin{array}{l}
\displaystyle
{\rm A}^+_{p}(\kappa)= 
\frac12\Big({\rm a}^+_{1}(\kappa)\ +  i\,{\rm a}^+_{2}(\kappa)\Big)
\\[2ex]
\displaystyle
{\rm A}^-_{m}(\kappa)= 
\frac12\Big({\rm a}^-_{1}(\kappa)\ - i\,{\rm a}^-_{2}(\kappa)\Big)
\\[2ex]
\displaystyle
{\rm A}_3(\kappa)=\frac12\Big(-\omega + 
 \kappa {\rm J}_{12} \Big)
\\[2ex]
\displaystyle
{\rm B}^+_{m}(\kappa)= 
\frac12\Big({\rm a}^+_{1}(\kappa) \ - i\,{\rm a}^+_{2}(\kappa)\Big)
\\[2ex]
\displaystyle
{\rm B}^-_{p}(\kappa)= 
\frac12\Big({\rm a}^-_{1}(\kappa) \ + i\,{\rm a}^-_{2}(\kappa,\omega)\Big)
\\[2ex]
\displaystyle
{\rm B}_3(\kappa)=\frac12\Big(  \omega + 
 \kappa {\rm J}_{12} \Big)
\end{array}
\end{equation}

Then, we obtain the following realization for these generators
\begin{equation}\label{abs}
\begin{array}{l}
\displaystyle
{\rm A}^+_p= \frac12  e^{i\phi}e^{-i\zeta \kappa} \left(- i p_\theta 
+\frac{  p_\phi}{{\rm T}_\kappa(\theta)} + p_\zeta\, {\rm T}_\kappa(\theta)\right)
\\[2.ex]
\displaystyle
{\rm A}^-_m= \frac12 e^{-i\phi}e^{i\zeta \kappa} \left(i p_\theta 
+\frac{ p_\phi}{{\rm T}_\kappa(\theta)} + p_\zeta\,{\rm T}_\kappa(\theta)\right)
\\[2.ex]
\displaystyle
{\rm A}_3 =\frac12\Big(- p_\zeta + 
  \kappa  p_\phi \Big)
\\[2ex]
\displaystyle
{\rm B}^+_m= \frac12 e^{-i\phi} e^{-i\zeta \kappa}\left(-i p_\theta 
-\frac{ p_\phi}{{\rm T}_\kappa(\theta)} + p_\zeta  {\rm T}_\kappa(\theta)\right) 
\\[2.ex]
\displaystyle
{\rm B}^-_p= \frac12 e^{i\phi} e^{i\zeta \kappa}\left(i p_\theta 
-\frac{  p_\phi}{{\rm T}_\kappa(\theta)} + p_\zeta \, {\rm T}_\kappa(\theta)\right) 
\\[2.ex]
\displaystyle
{\rm B}_3=\frac12\Big(p_\zeta +\, \kappa p_\phi \Big)
\end{array}
\end{equation}

\begin{figure}[h]
	\centering
\includegraphics[width= 8 cm]{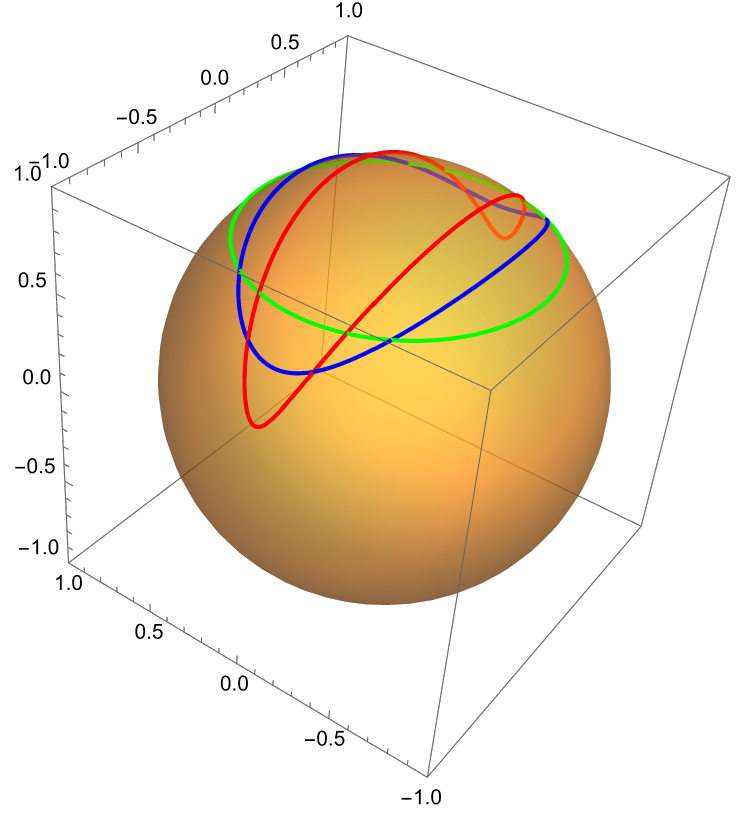}
\caption{\small Some trajectories on the sphere for different values
of the constants $\tilde {\rm Q}_{AA}, \tilde {\rm Q}_{BB}$ and
$\tilde {\rm Q}_{AB}$. The values of these plots are $E=8,12,40$ with $\ell=2$, $\omega=2$.
\label{sphere}
}
\end{figure}

\begin{figure}[h!]
	\centering
\includegraphics[width= 8 cm]{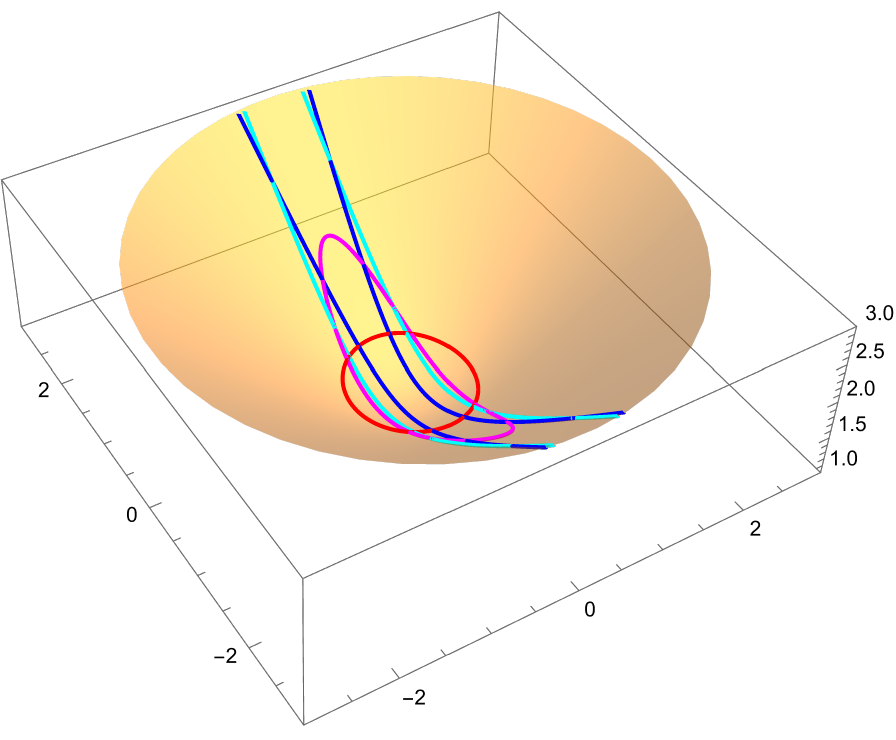}
\caption{\small Some trajectories on the hyperboloid for different values
of the constants $\tilde {\rm Q}_{AA}, \tilde {\rm Q}_{BB}$ and
$\tilde {\rm Q}_{AB}$. The values of these plots are $E=6,8,9,40$ with $\ell=3$, $\omega=1$.
\label{hyperbola}
}
\end{figure}

These functions close a $so_\kappa(4)$ Lie algebra.  Three independent symmetries are
\[
\tilde {\rm Q}_{AA}= A_+ A_-\,,\quad  
\tilde {\rm Q}_{BB}= B_+ B_-\,,\quad
\tilde {\rm Q}_{AB}= A_+ B_+ 
\]
The trajectories are obtained by fixing the values of the constants:
\begin{equation}\label{const}
\tilde {\rm Q}_{AA}=q_{a}^2,\quad \tilde {\rm Q}_{BB}=q_{b}^2,\quad \tilde {\rm Q}_{AB}=e^{i 2\phi_0}q_{a}q_{b}
\end{equation}
There are three independent real constants: $q_{a}$, $q_{b}$
and the angle $\phi_0$. The values of  $q_{a}$, $q_{b}$ are determined by the energy and momentum, which fix the form of the curves, while $\phi_0$ is for the orientation. 
We have the following relation
\[
2(\phi-\phi_0)= \arccos\frac{\left(\frac{\ell}{{\rm T}_\kappa(\theta)} + \omega\, {\rm T}_\kappa(\theta)\right)}{q_a}
+
\arccos\frac{\left(\frac{ - \ell}{{\rm T}_\kappa(\theta)} + \omega\, {\rm T}_\kappa(\theta)\right)}{q_b}
\]
After some computations the trajectories take the algebraic form
\[
\cos2(\phi-\phi_0)=\frac{q_a^2+ q_b^2 
+2\ell^2/{\rm T}^2_\kappa(\theta)}{q_a q_b}
\]
where $q_a^2-q_b^2= 4\omega\ell$.
In the case of the sphere all the curves are closed and restricted to one of the hemispheres; in the hyperboloid, the curves may be closed, open limiting curves  or open type orbits, see Figs.~\ref{sphere} and \ref{hyperbola}. For an exhaustive discussion of the orbits, see ref.~\cite{carinena08}.

\section{Conclusions}

Along this work we have obtained, for the curved HO, the analog of the creation/annihilation operators of the flat HO. In this way we have arrived to a general form of the FD tensor.
We have worked out the case of two dimensions, to show the main features of the construction in a simple way, and in order to keep the paper within a reasonable extension.
We plan to supply the results of the 3D HO in a future publication.  

We have included positive or negative curvatures through the parameter $\kappa$,  such that in the limit $\kappa \to 0$ we obtain the well known flat operators. Therefore, we can appreciate how the operators in curved space are defined by a natural modification of the flat space formulas. Nevertheless,  there  appear important differences in curved spaces. The most important is the algebra closed by these basic ``curved'' operators, which (in 2D) we have identified as a direct sum $so_\kappa(3)\oplus so_\kappa(3)$. We have obtained two sets of basic operators that we called parallel and spherical basis, where in the limit they produce the Cartesian or the polar basis of the flat 2D HO. The spherical basis proved to be the most useful in curved HO in order to find the representation space in the quantum case or the orbits in the classical case.

Although the degeneracy is the same for curved/flat systems, the eigenspaces belong to different representations of the involved algebra for the curved systems, while in the flat HO all the eigenfunctions belong to the same representation.  

Finally we remark that our construction, valid for quantum or classical systems,  is based in the factorization method. Therefore, we have shown through one more example (other recent examples are given in \cite{salamanca23,kuru17,latini16}) how this method can be quite fruitful when it is applied to superintegrable systems.

\section*{Acknowledgments}

The authors would like to acknowledge the support of the QCAYLE project, funded by the European Union--NextGenerationEU, PID2020-113406GB-I0 project funded by the MCIN of Spain and the contribution of the European Cooperation in Science and Technology COST Action CA23130.
\c{S}.~K. thanks Ankara University and the warm hospitality of the Department of Theoretical Physics of the University of Valladolid, where part of this work has been carried out, and to the support of its GIR of Mathematical Physics.

\end{document}